\documentclass[english,floatfix,showpacs,notitlepage]{revtex4-1}
\usepackage[T1]{fontenc}
\usepackage[latin9]{inputenc}
\usepackage{amsmath}
\usepackage{amssymb}
\usepackage{graphicx}

\makeatletter

\usepackage{babel}

\begin{document}

\title{Renormalization of Discrete-Time Quantum Walks with non-Grover Coins}

\author{Stefan Boettcher and Joshua L. Pughe-Sanford }

\affiliation{Department of Physics, Emory University, Atlanta, GA 30322; USA}
\begin{abstract}
We present an in-depth analytic study of discrete-time quantum walks
driven by a non-reflective coin. Specifically, we compare the properties of the widely-used
Grover coin ${\cal C}_{G}$ that is unitary and reflective (${\cal C}_{G}^{2}=\mathbb{I}$)
with those of a $3\times3$ ``rotational'' coin ${\cal C}_{60}$
that is unitary but non-reflective (${\cal C}_{60}^{2}\not=\mathbb{I}$)
and satisfies instead ${\cal C}_{60}^{6}=\mathbb{I}$, which corresponds
to a rotation by $60^{\circ}$. While such a modification apparently
changes the real-space renormalization group (RG) treatment, we show
that nonetheless this non-reflective quantum walk remains in the same
universality class as the Grover walk. We first demonstrate the procedure
with ${\cal C}_{60}$ for a 3-state quantum walk on a one-dimensional
(\emph{1d}) line, where we can solve the RG-recursions in closed form,
in the process providing exact solutions for some difficult non-linear
recursions. Then, we apply the procedure to a quantum walk on a dual
Sierpinski gasket (DSG), for which we reproduce ultimately the same
results found for ${\cal C}_{G}$, further demonstrating the robustness
of the universality class.
\end{abstract}
\maketitle
\section{Introduction\label{sec:Intro}}

Recent studies of quantum walks~\cite{Aharonov93,Meyer96,AAKV01,Kempe03,PortugalBook} using the real-space renormalization group (RG)~\cite{Boettcher16,Boettcher17a} have revealed the rich analytic structure of this quantum extension of classical random walks~\cite{Shlesinger84,Weiss94,Hughes96,Metzler04}, in which the stochastic operator in the master equation describing the dynamics is replaced with a unitary operator. Profound, new insights can be obtained, for example, for the asymptotic behavior of quantum search algorithms~\cite{Boettcher17b}. For either type of walk, it is fundamental to determine the spreading behavior of the probability density
function (PDF) $\rho\left(\vec{x},t\right)$, which predicts the likelihood to detect a walk at time
$t$ at site of distance $x=\left|\vec{x}\right|$ after starting
at the origin. Asymptotically,
the PDF obeys the scaling collapse with the scaling variable $x/t^{1/d_{w}}$,
\begin{equation}
\rho\left(\vec{x},t\right)\sim t^{-\frac{d_{f}}{d_{w}}}f\left(x/t^{\frac{1}{d_{w}}}\right),\label{eq:collapse}
\end{equation}
where $d_{w}$ is the walk-dimension and $d_{f}$ is the fractal dimension
of the network~\cite{Havlin87}. This scaling
impacts many important observables such as the mean-square
displacement, $\left\langle x^{2}\right\rangle _{t}\sim t^{2/d_{w}}$~\cite{SWN}.

One purpose of the RG is the exploration of universality classes in the dynamics of physical 
systems~\cite{Plischke94}. It reveals how internal symmetries - and their breaking -  may affect the 
large-scale behavior of the system. These methods have been adapted to explore the asymptotic 
scaling of random walks~\cite{Havlin87,Hughes96,Redner01} and quantum 
walks~\cite{Boettcher13a, QWNComms13}. We have derived exact RG-flow equations for 
discrete-time ("coined") quantum walks on a number
of complex networks~\cite{QWNComms13,Boettcher14b} and argued that the walk dimension $d_{w}$
in Eq.~(\ref{eq:collapse}) for a quantum walk with a Grover coin
${\cal C}_{G}$ always is \emph{half} of that for the corresponding
random walk, $d_{w}^{Q}=\frac{1}{2}\,d_{w}^{R}$, first based on numerical evidence~\cite{Boettcher14b} and later substantiated by analytic results~\cite{Boettcher16,Boettcher17a}. Furthermore, this RG can be used to determine that the computational
complexity of a Grover search algorithm~\cite{Gro97a} merely depends on the spectral
dimension $d_{s}$ of the network that represents the database on
which the quantum search is conducted~\cite{Boettcher17b}. It was
not obvious, however, whether such a result would survive the breaking
of certain symmetries such as the reflectivity of the Grover coin,
${\cal C}_{G}^{2}=\mathbb{I}$, that is a remnant of Grover's original
conception~\cite{Gro97a} of a quantum walk on a mean-field, all-to-all
network. Here, we present a first exploration of the extent of the
universality class by using a non-reflective coin ${\cal C}_{60}$
that is equally unitary but non-reflective (${\cal C}_{G}^{2}\not=\mathbb{I}$)
and instead satisfies ${\cal C}_{G}^{6}=\mathbb{I}$, amounting to
a $60^{\circ}$-degree rotation in the coin-space of the quantum walker.
We discuss in detail the seemingly significant changes for the RG
the new coin entails by studying the solvable case of a walk on the
\emph{1d}-line before we extend those insights to the nontrivial case
of a quantum walker on a dual Sierpinski gasket (DSG). Our findings,
however, suggest that the  scaling behavior is ultimately
unaffected by the change in coin. Thus, the universality class of
the Grover walk appears to be  robust.

Along the way, we also address several technical issues regarding
the RG. The RG for classical random walks is straightforward~\cite{Redner01}:
The Laplace-poles for normalized hopping parameters as well as observables
alike impinge on a fixed point at $z=1$ along the real $z$-axis
to connect asymptotic scaling in space and time, as expressed by the
walk dimension $d_{w}$ defined in Eq.~(\ref{eq:collapse}). In the
case of a quantum walk, poles accumulated in wedges, either along
the unit-circle in the complex-$z$ plane for the hopping parameters,
or constrained on the circle for the unitary observables, leading
to a distinct interpretation~\cite{Boettcher16,Boettcher17a}. While
for the reflective Grover coins ${\cal C}_{G}$ these poles still
impinge on either or both of $z=\pm1$, the use of the non-reflective
coin ${\cal C}_{60}$ affords us the opportunity to explore the interpretation
of situations where such poles impinge on fixed points at non-trivial
$z_{0}$ on the unit-circle in model problems that are exactly solvable,
i.e., the \emph{1d}-line. We then test those interpretations for the
DSG, where no such solution exists. We find that precise knowledge
of $z_{0}$ is not required for the RG analysis, although isolated
exceptional points seem to exist that must be avoided.

This paper is organized as follows: In Sec.~\ref{sec:Quantum-Master},
we discuss the basic features of the RG, and compare the exact results
for the quantum walk on the \emph{1d}-line for both coins. In 
Sec.~\ref{sec:Quantum-Walk-DSG}, we apply the same scheme to analyze the
non-trivial case of quantum walks on DSG. In Sec.~\ref{sec:Discussion},
we conclude with a summary and discussion of our results.

\section{Quantum Walk on the Line\label{sec:Quantum-Master}}

The time evolution of a quantum walk is governed by the discrete-time
master equation~\cite{PortugalBook} 
\begin{equation}
\left|\Psi_{t+1}\right\rangle ={\cal U}\left|\Psi_{t}\right\rangle \label{eq:MasterEq}
\end{equation}
with unitary propagator ${\cal U}$. With $\psi_{x,t}=\left\langle x|\Psi_{t}\right\rangle $
in the discrete $N$-dimensional site-basis $\left|x\right\rangle $,
the probability density function is given by $\rho\left(x,t\right)=\left|\psi_{x,t}^{2}\right|$.
In this basis, the propagator can be represented as an $N\times N$
matrix ${\cal U}_{x,y}=\left\langle x\left|{\cal U}\right|y\right\rangle $
with operator-valued entries that describe the transitions between
neighboring sites (``hopping matrices''). We can study the long-time
asymptotics via a discrete Laplace transform, 
\begin{equation}
\overline{\psi}_{x}\left(z\right)={\textstyle \sum_{t=0}^{\infty}}\psi_{x,t}z^{t},\label{eq:LaplaceT}
\end{equation}
as $\left|z\right|\to1$ implies the limit $t\to\infty$, in a manner
that we will have to specify. Then, Eq.~(\ref{eq:MasterEq}) becomes
\begin{equation}
\overline{\psi}_{x}=z{\cal U}_{x,y}\overline{\psi}_{y}+\psi_{x,t=0}.\label{eq:z_master}
\end{equation}

We review the time evolution of quantum walks on the \emph{1d}-line
for which the propagator in Eq.\ (\ref{eq:MasterEq}) is 
\begin{equation}
{\cal U}={\textstyle \sum_{x}}\left\{ A\left|x+1\right\rangle \left\langle x\right|+B\left|x-1\right\rangle \left\langle x\right|+M\left|x\right\rangle \left\langle x\right|\right\} \label{eq:propagator}
\end{equation}
for nearest-neighbor transitions. The norm of $\rho$ for quantum
walks demands unitary propagation, $\mathbb{I}={\cal U}^{\dagger}{\cal U}$,
which then imposes the conditions $\mathbb{I}_{r}=A^{\dagger}A+B^{\dagger}B+M^{\dagger}M$
and $0=A^{\dagger}M+M^{\dagger}B=A^{\dagger}B$, consistent with $A+B+M$
being unitary. As there are no non-trivial choices for scalar solutions
of the unitarity constraints, one extends Hilbert space to include
an internal degree of freedom for the wave-function, the so-called
coin-space, similar to giving the walker a spin. Often, one defines
the propagator then as ${\cal U}={\cal S}\left(\mathbb{I}_{N}\otimes{\cal C}\right)$,
with ``coin'' ${\cal C}$ and ``shift'' ${\cal S}$. The
quantum walk entangles spatial and coin-degrees of freedom. First,
the coin mixes all components of $\psi_{x,t}$, then the shift-matrices
$S^{\left\{ A,B,M\right\} }$ facilitate the subsequent transition
to neighboring sites or the same site, respectively. We obtain thus the hopping
matrices $A=S^{A}{\cal C}$, $B=S^{B}{\cal C}$, and $M=S^{M}{\cal C}$
with $S^{A}+S^{B}+S^{M}=\mathbb{I}$. Because the most general 2-dimensional
unitary coin is always also reflective (aside from a trivial rotation)~\cite{Boettcher13a}, 
to allow for a non-reflective coin requires
at least a 3-dimensional coin-space, for which it is convenient to
choose
\begin{equation}
S^{A}=\left[\begin{array}{ccc}
1 & 0 & 0\\
0 & 0 & 0\\
0 & 0 & 0
\end{array}\right],\quad S^{B}=\left[\begin{array}{ccc}
0 & 0 & 0\\
0 & 1 & 0\\
0 & 0 & 0
\end{array}\right],\quad S^{M}=\left[\begin{array}{ccc}
0 & 0 & 0\\
0 & 0 & 0\\
0 & 0 & 1
\end{array}\right].\label{eq:Shift}
\end{equation}

\begin{figure}
\hfill{}\includegraphics[clip,width=0.9\columnwidth]{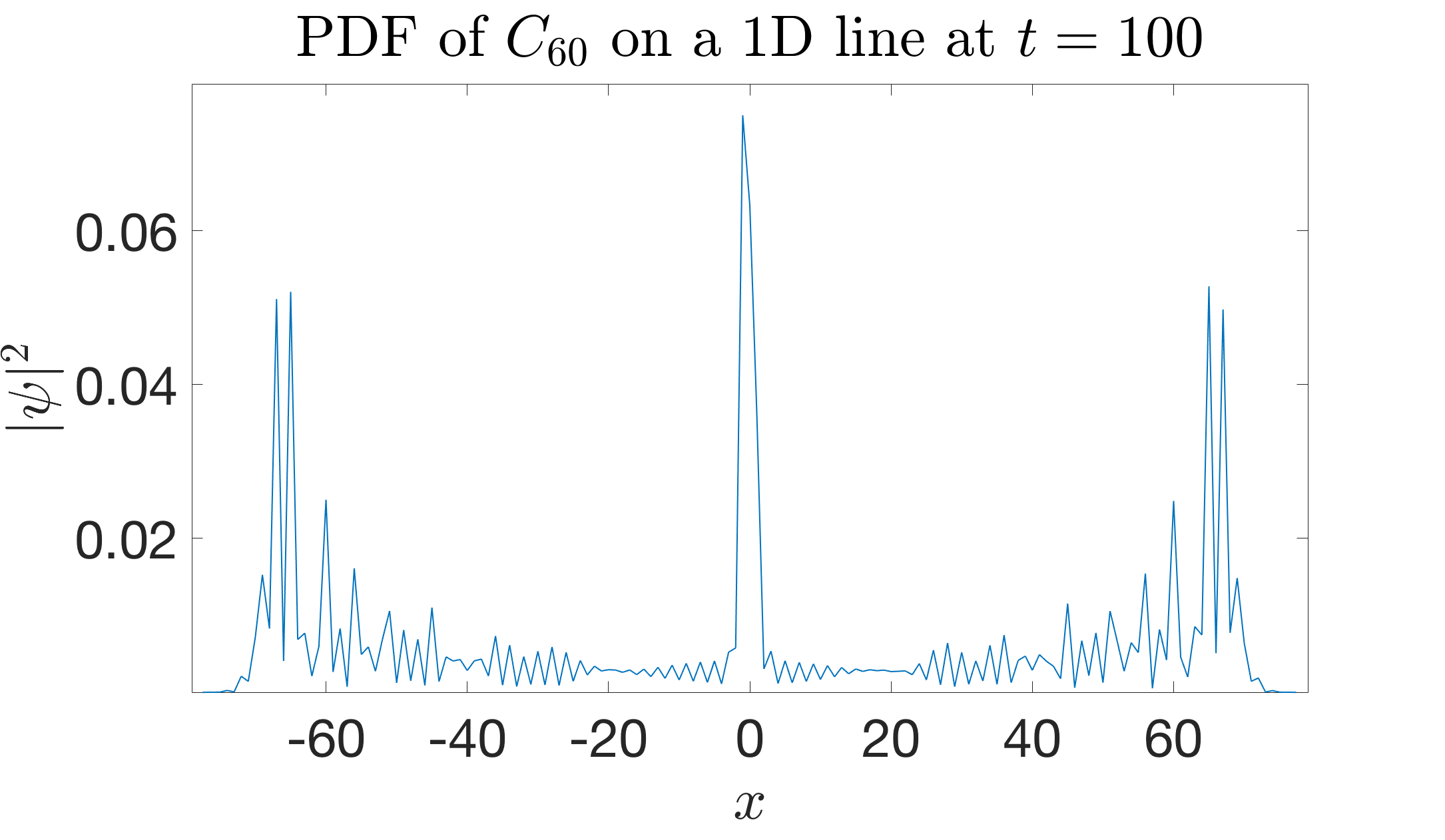}\hfill{}

\caption{\label{fig:1d3statenonGroversnapshot}Snapshot of the probability
density function $\rho\left(x,t\right)=\left|\psi_{x,t}^{2}\right|$
after $t=100$ updates using ${\cal C}_{60}$ in Eq.~(\ref{eq:nonGroverC})
for a quantum walk starting from the origin. As for the equivalent
walk with the Grover coin in Eq.~(\ref{eq:GroverC}), see Refs.~\cite{Inui05,Falkner14a},
this walk exhibits a permanently localized component near its origin,
while the remaining weight of the wave-function spreads ballistically
in both directions along the lattice. }
\end{figure}
Such a 3-state quantum walk on the 1d-line has been investigated previously
for a $3\times3$ Grover coin~\cite{Inui05,Falkner14a}, 
\begin{equation}
{\cal C}_{G}=\frac{1}{3}{\textstyle \left[\begin{array}{rrr}
-1 & 2 & 2\\
2 & -1 & 2\\
2 & 2 & -1
\end{array}\right]}.\label{eq:GroverC}
\end{equation}
While the 3-dimensional coin attains the same asymptotic scaling as
the more widely studied 2-dimensional case, it adds the interesting
new aspect of localization to the dynamics~\cite{Inui05,Falkner14a},
where the walker becomes eternally confined near its starting site
with a finite probability. In the following, we will investigate the
effects of a non-reflective coin,
\begin{equation}
{\cal C}_{60}=\frac{1}{3}{\textstyle \left[\begin{array}{rrr}
2 & 2 & -1\\
-1 & 2 & 2\\
2 & -1 & 2
\end{array}\right]}.\label{eq:nonGroverC}
\end{equation}
As the geometry is  simple, it is not too surprising that we
find qualitatively the same phenomenology for both  coins,
as shown in Fig.~\ref{fig:1d3statenonGroversnapshot}. Thus, we will
not bother with the canonical Fourier solution of the problem that
can be found elsewhere~\cite{Inui05,Falkner14a}. Instead, we will
immediately proceed to treat this problem as an exactly solvable example
for the RG from which we can glean  insights on how to treat
more complicated non-Grover quantum walks with RG.

\begin{figure}

\hfill{} \includegraphics[bb=5bp 220bp 500bp 590bp,clip,width=.3\columnwidth]{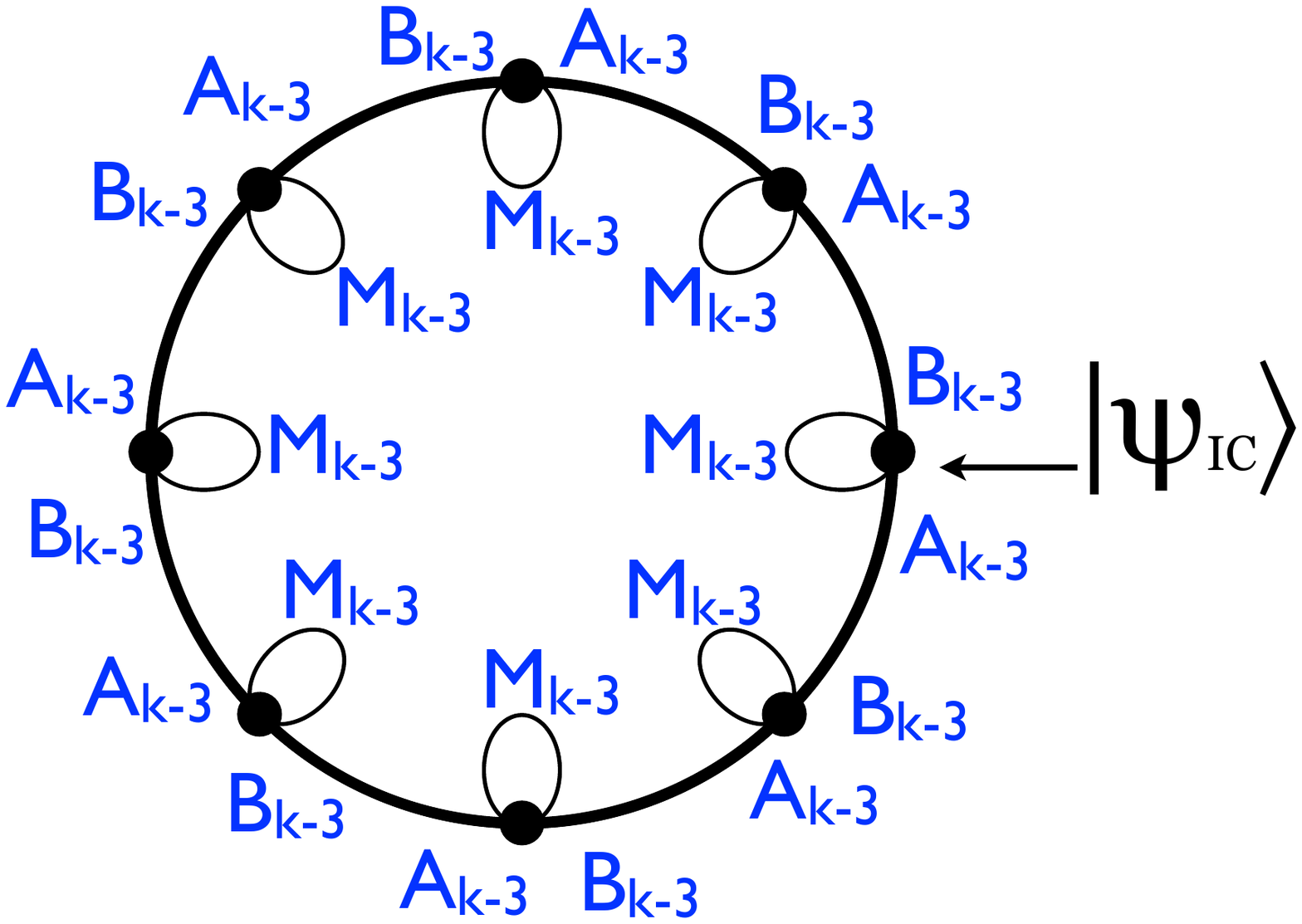}
\hfill{} \includegraphics[bb=5bp 220bp 500bp 590bp,clip,width=.3\columnwidth]{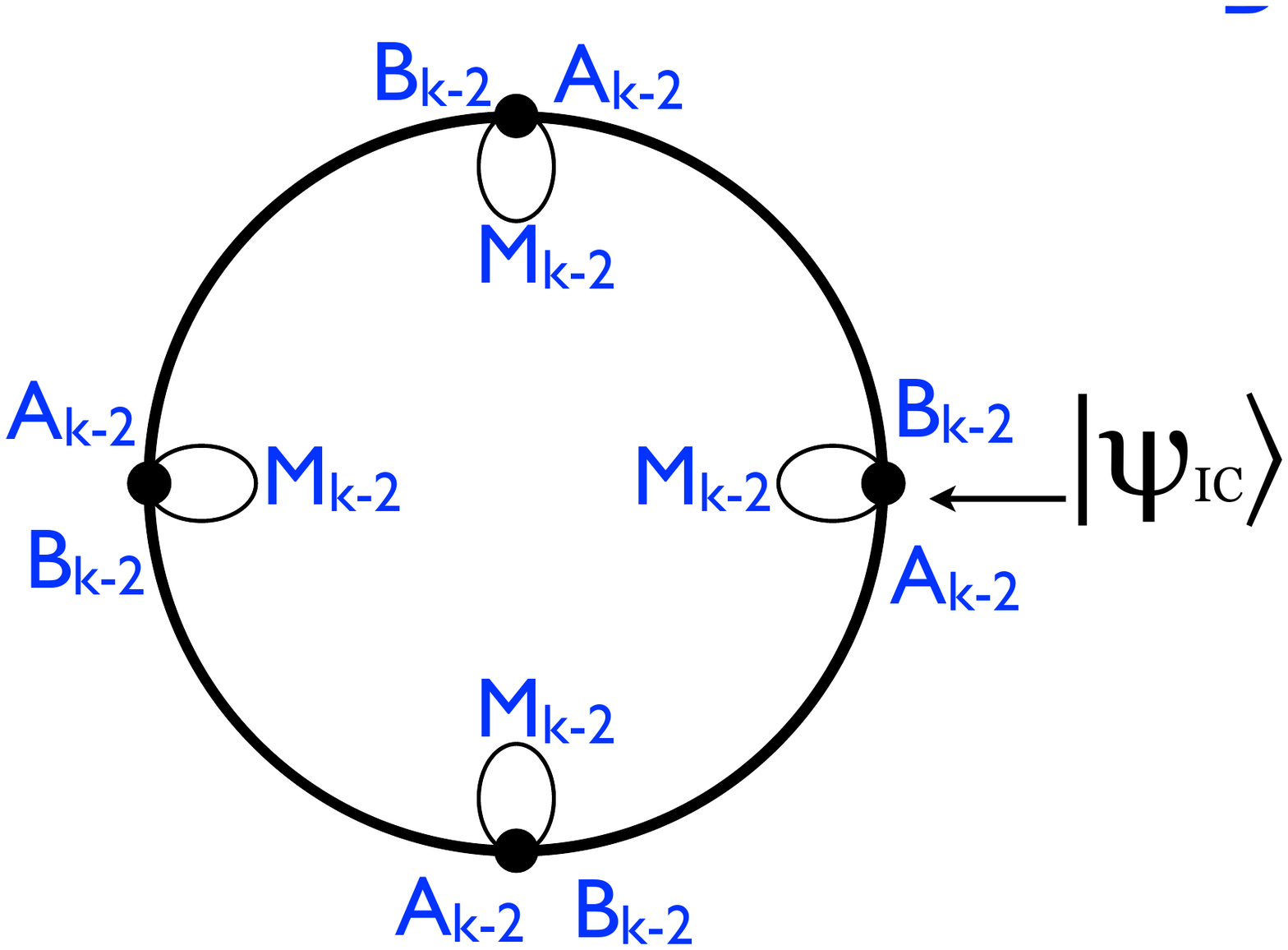}
\hfill{} \includegraphics[bb=5bp 220bp 500bp 590bp,clip,width=.3\columnwidth]{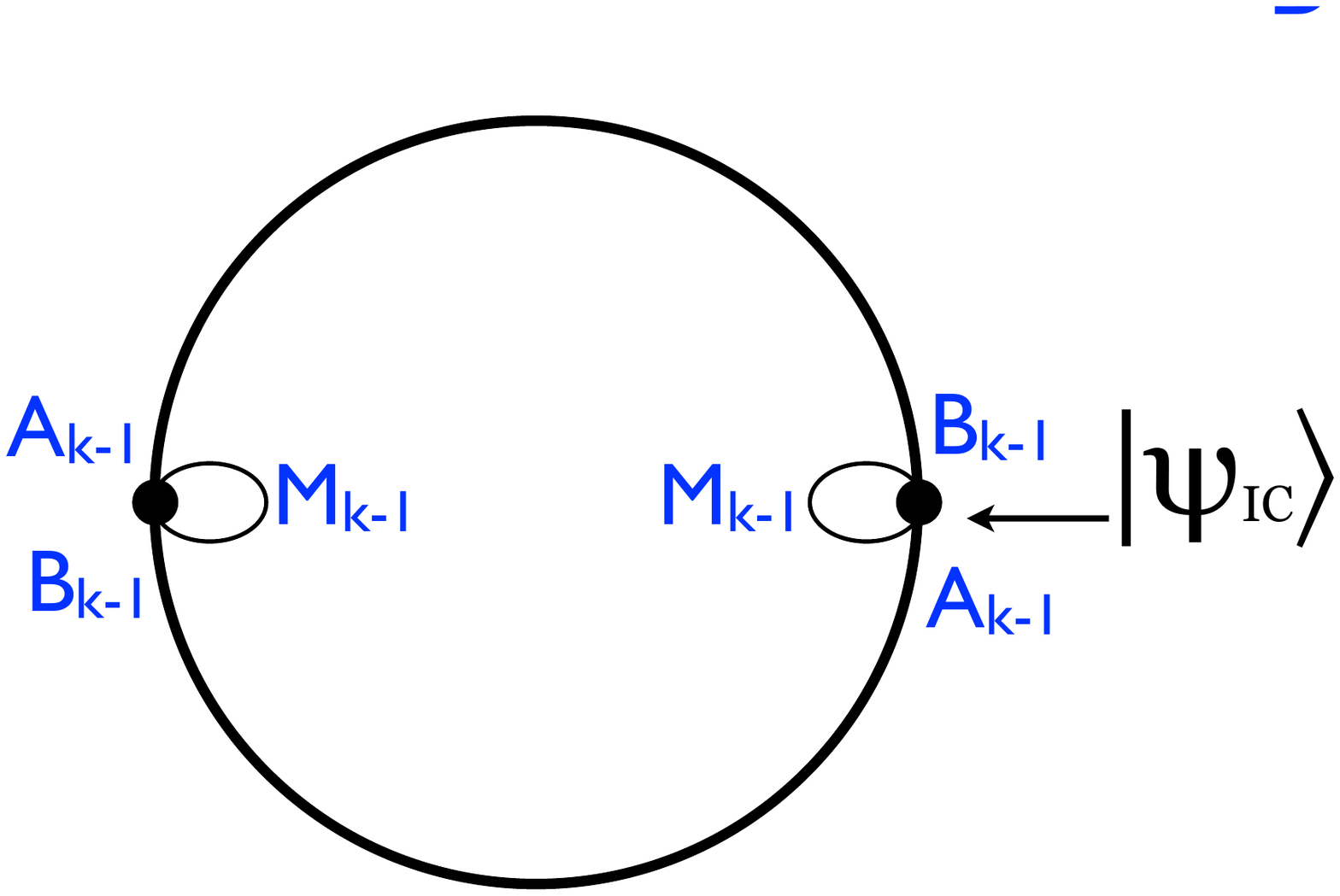}\hfill{}

\caption{\label{fig:1dQWRG}Depiction of the final three RG-steps for a
simple \emph{1d}-line.}
\end{figure}

\subsection{RG for the Quantum Walk on the Line\label{sec:RG-for-the}}

The transformed master equation (\ref{eq:z_master}) with ${\cal U}$
in Eq.~(\ref{eq:propagator}) becomes $\overline{\psi}_{x}=zM\overline{\psi}_{x}+zA\overline{\psi}_{x-1}+zB\overline{\psi}_{x+1}$,
which is the starting point for the RG. For simplicity, we consider
the walk problem here with initial conditions (IC) localized at the
origin, $\psi_{x,t=0}=\delta_{x,0}\psi_{IC}$. As depicted in Fig.~\ref{fig:1dQWRG},
we recursively eliminate $\overline{\psi}_{x}$ for all sites for
which $x$ is an odd number, then set $x\to x/2$ and repeat, step-by-step
for $k=0,1,2,\ldots$. Starting at $k=0$ with the ``raw'' hopping
coefficients $A_{0}=zA$, $B_{0}=zB$, and $M_{0}=zM$, after each
step, the master equation becomes \emph{self-similar} in form when
redefining the renormalized hopping coefficients $A_{k}$, $B_{k}$,
$M_{k}$. E.g.,  near any even site $x$ at step $k$ we
have~\cite{Boettcher13a}: 
\begin{eqnarray}
\overline{\psi}_{x-1} & = & M_{k}\overline{\psi}_{x-1}+A_{k}\overline{\psi}_{x-2}+B_{k}\overline{\psi}_{x},\nonumber \\
\overline{\psi}_{x} & = & M_{k}\overline{\psi}_{x}+A_{k}\overline{\psi}_{x-1}+B_{k}\overline{\psi}_{x+1}+\delta_{x,0}\psi_{IC},\label{eq:1dPRWmass-master}\\
\overline{\psi}_{x+1} & = & M_{k}\overline{\psi}_{x+1}+A_{k}\overline{\psi}_{x}+B_{k}\overline{\psi}_{x+2}.\nonumber 
\end{eqnarray}
Solving this \emph{linear} system for site $x$ yields $\overline{\psi}_{x}=M_{k+1}\overline{\psi}_{x}+A_{k+1}\overline{\psi}_{x-2}+B_{k+1}\overline{\psi}_{x+2}$
with RG ``flow'' 
\begin{eqnarray}
A_{k+1} & = & A_{k}\left(\mathbb{I}-M_{k}\right)^{-1}A_{k},\qquad B_{k+1}=B_{k}\left(\mathbb{I}-M_{k}\right)^{-1}B_{k},\label{eq:recur1dPRWmass}\\
M_{k+1} & = & M_{k}+A_{k}\left(\mathbb{I}-M_{k}\right)^{-1}B_{k}+B_{k}\left(\mathbb{I}-M_{k}\right)^{-1}A_{k}.\nonumber 
\end{eqnarray}

\subsection{RG-Analysis for the Grover Coin\label{subsec:RG-for-Grover}}

For the quantum walk with the Grover coin ${\cal C}_{G}$ in Eq.~(\ref{eq:nonGroverC}),
we evolve Eqs.\ (\ref{eq:recur1dPRWmass}) for a few iterations from
its unrenormalized values~\cite{Boettcher13a}. Already after
the first iteration, a recurring pattern emerges that suggest the
Ansatz $A_{k}=a_{k}S^{A}{\cal C}_{G}$, $B_{k}=a_{k}S^{B}{\cal C}_{G}$,
and
\begin{equation}
M_{k}=\left[\begin{array}{ccc}
0 & b_{k} & 0\\
b_{k} & 0 & 0\\
0 & 0 & z
\end{array}\right]{\cal C}_{G},\label{eq:MkGrover}
\end{equation}
where at $k=1$ the RG-flow gets initiated with
\begin{eqnarray}
a_{k=1}=\frac{z^{2}\left(z-1\right)}{z+3}, & \quad & b_{k=1}=\frac{2z^{2}\left(z+1\right)}{z+3}.\label{eq:RGinit1dGrover}
\end{eqnarray}
For this scalar parametrization with $a_{k}$ and $b_{k}$, the RG-flow
(\ref{eq:recur1dPRWmass}) closes after each iteration for 
\begin{eqnarray}
a_{k+1} & = & \frac{a_{k}^{2}\left(z-1\right)}{\left(b_{k}-1\right)\left[\left(3z+1\right)b_{k}-z-3\right]},
\label{eq:1dRGflowGrover}\\
b_{k+1} & = & b_{k}+\frac{a_{k}^{2}\left[\left(3z+1\right)b_{k}-2z-2\right]}{\left(b_{k}-1\right)\left[\left(3z+1\right)b_{k}-z-3\right]}\nonumber
\end{eqnarray}
These recursions have a non-trivial fixed point at $\left(a_{\infty},b_{\infty}\right)=\left(\frac{1-z}{1+3z},\frac{2\left(1+z\right)}{1+3z}\right)$,
yet, its Jacobian $J_{\infty}=\left.\frac{\partial\left(a_{k+1},b_{k+1}\right)}{\partial\left(a_{k},b_{k}\right)}\right|_{k\to\infty}$
is, in fact, $z$-independent and has a degenerate eigenvalue $\lambda_{1,2}=2$.
Note that we never had to specify a choice for $z$ as the location
of a fixed point to arrive at this result. However, numerical iteration
of the RG-flow in Eq.~(\ref{eq:1dRGflowGrover}) from the initial
conditions in Eq.~(\ref{eq:RGinit1dGrover}) clearly shows that fixed
points are only attained for very specific choices of $z$. Those
values of $z$ can be obtained by equating $\left(a_{\infty},b_{\infty}\right)=\left(a_{k},b_{k}\right)$,
which provides an over-constraint condition on solving for $z$. For
instance, the initial conditions in Eq.~(\ref{eq:RGinit1dGrover})
merely yield $z=\pm1$ as fixed points for $k=1$, $k=2$ adds $z=-\left(2\pm i\sqrt{5}\right)/3$,
etc. We will return to this issue later.

Previously~\cite{Boettcher16}, it was observed that is not sufficient
to consider the hopping parameters $\left(a_{k},b_{k}\right)$ alone.
To study the true scaling behavior, actual observables need to be
examined; although observables are functionals of the hopping parameters,
they are constrained to be unitary, unlike the hopping parameters.
As in the classical analysis of a random walk~\cite{Redner01}, there
is an intimate connection between fixed points and the poles in the
complex-$z$ plane. The evolution under rescaling size ($k$) of those
poles nearest to the fixed point yields the rescaling in time. While
those poles move both radially as well as tangentially along the unit-circle
for hopping parameters, they are constrained to moving only tangentially
strictly on the circle for a unitary observable, leading to the walk
dimension
\begin{equation}
d_{w}^{Q}=\log_{2}\sqrt{\lambda_{1}\lambda_{2}}.\label{eq:dwQ}
\end{equation}
Hence, with $\lambda_{1}=\lambda_{2}=2$, we obtain $d_{w}^{Q}=1$,
as expected for a quantum walk on the \emph{1d}-line. We proceed to
investigate such an observable more closely. 

As shown in Fig.~\ref{fig:1dQWRG}, the 1d-loop of $N=2^{k}$ sites after $k-1$ RG-steps reduces to merely
two sites:
\begin{eqnarray}
\overline{\psi}_{0} & = & M_{k-1}\overline{\psi}_{0}+\left(A_{k-1}+B_{k-1}\right)\overline{\psi}_{\frac{N}{2}}+\psi_{IC},\nonumber\\
\overline{\psi}_{\frac{N}{2}} & = & M_{k-1}\overline{\psi}_{\frac{N}{2}}+\left(A_{k-1}+B_{k-1}\right)\overline{\psi}_{0}.\label{FinalRG}
\end{eqnarray}
Eliminating also the site $\overline{\psi}_{\frac{N}{2}}$, and utilizing
the RG-recursions in Eq.~(\ref{eq:recur1dPRWmass}), we get the amplitude
at the origin $x=0$ of a quantum walk, $\overline{\psi}_{0}=X_{k}\psi_{IC}$,
for the \emph{1d}-line: 
\begin{equation}
X_{k}=\left[\mathbb{I}-\left(A_{k}+B_{k}+M_{k}\right)\right]^{-1}.\label{eq:XkDef}
\end{equation}
It is now straightforward from the expressions near Eq.~(\ref{eq:MkGrover})
to express the observable $X_{k}$ in terms of the hopping parameters
$\left(a_{k},b_{k}\right)$:
\begin{equation}
X_{k}=\frac{\left[\begin{array}{ccc}
\frac{-z-3+a_{k}\left(z-1\right)+2b_{k}\left(z+1\right)}{1+a_{k}-b_{k}} & \frac{-2a_{k}\left(z+1\right)-b_{k}\left(z-1\right)}{1+a_{k}-b_{k}} & -2a_{k}-2b_{k}\\
\frac{-2a_{k}\left(z+1\right)-b_{k}\left(z-1\right)}{1+a_{k}-b_{k}} & \frac{-z-3+a_{k}\left(z-1\right)+2b_{k}\left(z+1\right)}{1+a_{k}-b_{k}} & -2a_{k}-2b_{k}\\
-2z & -2z & a_{k}+b_{k}-3
\end{array}\right]}{\left(a_{k}+b_{k}\right)\left(3z+1\right)-\left(z+3\right)}.\label{eq:Xkakbk}
\end{equation}
It is further fruitful to first solve the RG-flow in Eq.~(\ref{eq:1dRGflowGrover})
in closed form. A simple rational Ansatz, amazingly, yields the \emph{exact}
solution
\begin{eqnarray}
a_{k} & = & \frac{\left(u-1\right)\left(z-1\right)\left(3uz+u-z-3\right)\xi^{2^{k-1}}}{\left(z-1\right)^{2}-\left(3uz+u-2z-2\right)^{2}\xi^{2^{k}}},\label{eq:Solk1dGrover}\\
b_{k} & = & \frac{u\left(z-1\right)^{2}-\left(3uz+u-2z-2\right)\left(2uz+2u-z-3\right)\xi^{2^{k}}}{\left(z-1\right)^{2}-\left(3uz+u-2z-2\right)^{2}\xi^{2^{k}}},\nonumber 
\end{eqnarray}
where the otherwise free constants $u$ and $\xi$ are fixed by the
initial conditions at $k=1$ in Eq.~(\ref{eq:RGinit1dGrover}):
\begin{eqnarray}
\xi & = & 16+9\cos\left(2\omega\right)+24\cos\omega-\cos\frac{\omega}{2}\left(6\cos\omega+4\right)\sqrt{18\cos\omega+6},\nonumber \\
u & = & e^{i\omega}\left[\frac{3}{2}\cos\omega-\frac{1}{2}+\frac{1}{2}\cos\frac{\omega}{2}\sqrt{18\cos\omega+6}\right],\label{eq:XiUGrover}
\end{eqnarray}
using $z=e^{i\omega}$. Note that $\xi$ is complex (with $\left|\xi\right|=1$)
only for $\omega=\arg z\in I$ on the interval $I=\left(\pi-\arctan2\sqrt{2},\pi+\arctan2\sqrt{2}\right)$.
An infinity of solutions for $\omega$ of $\xi^{2^{j}}=1$ for any
integer $j\geq0$ form a (likely dense) set of fixed points of the
RG-flow on $I$, because they leave Eq.~(\ref{eq:Solk1dGrover}) invariant
for all $j\leq k\leq\infty$.

It is instructive to define $\xi=e^{i\nu}$ and $u=\left[2\left(z+1\right)+ie^{i\sigma}\left(z-1\right)\right]/\left(3z+1\right)$
to get
\begin{eqnarray}
\cos\nu & = & 9\cos\left(2\omega\right)+24\cos\left(\omega\right)+16,\nonumber \\
\sin\sigma & = & -\frac{9}{4}z^{2}-\frac{3}{2}z-\frac{5}{4}.\label{eq:SigNu}
\end{eqnarray}
Then, the solution of the RG-flow given in Eq.~(\ref{eq:Solk1dGrover})
simplifies to
\begin{eqnarray}
a_{k} & = & \frac{\left(1-z\right)\cos\sigma}{\left(3z+1\right)\cos\left(2^{k}\nu+\sigma\right)},\label{eq:akbkinsignu}\\
b_{k} & = & \frac{2\left(z+1\right)}{3z+1}+\frac{\left(z-1\right)\sin\left(2^{k}\nu\right)}{\left(3z+1\right)\cos\left(2^{k}\nu+\sigma\right)}.\nonumber 
\end{eqnarray}
From Eq.~(\ref{eq:Xkakbk}) we thus find
\begin{eqnarray}
X_{k} & = & \frac{\chi}{\left(z-1\right)\sin\left(2^{k-1}\nu\right)},\label{eq:XinChi}
\end{eqnarray}
with a regular numerator matrix $\chi$ that does not have any $k$-dependent
poles. 

\begin{figure*}
\hfill{}\includegraphics[bb=0bp 0bp 1000bp 1000bp,clip,width=0.48\columnwidth]{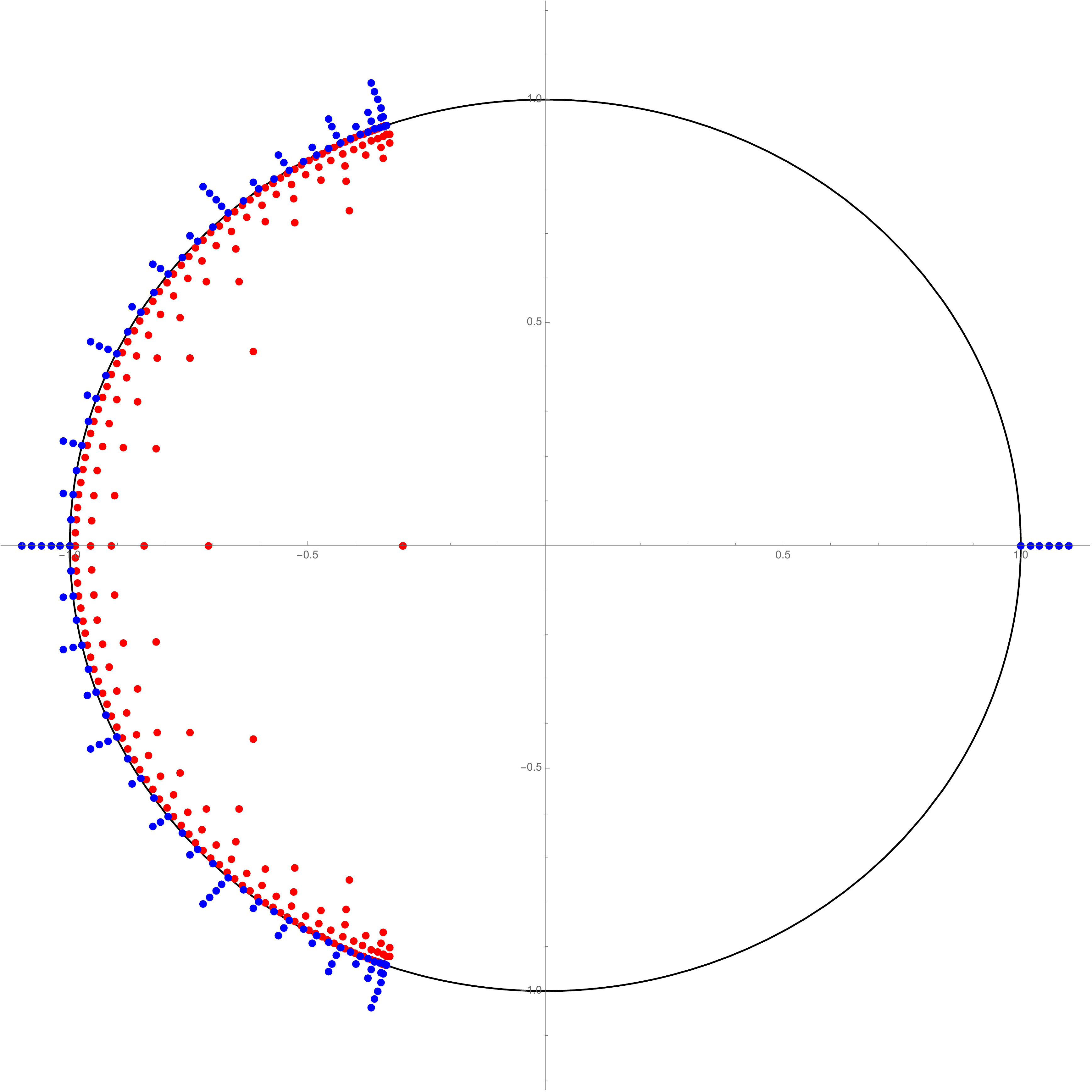}\hfill{}\includegraphics[bb=0bp 0bp 1000bp 1000bp,clip,width=0.48\columnwidth]{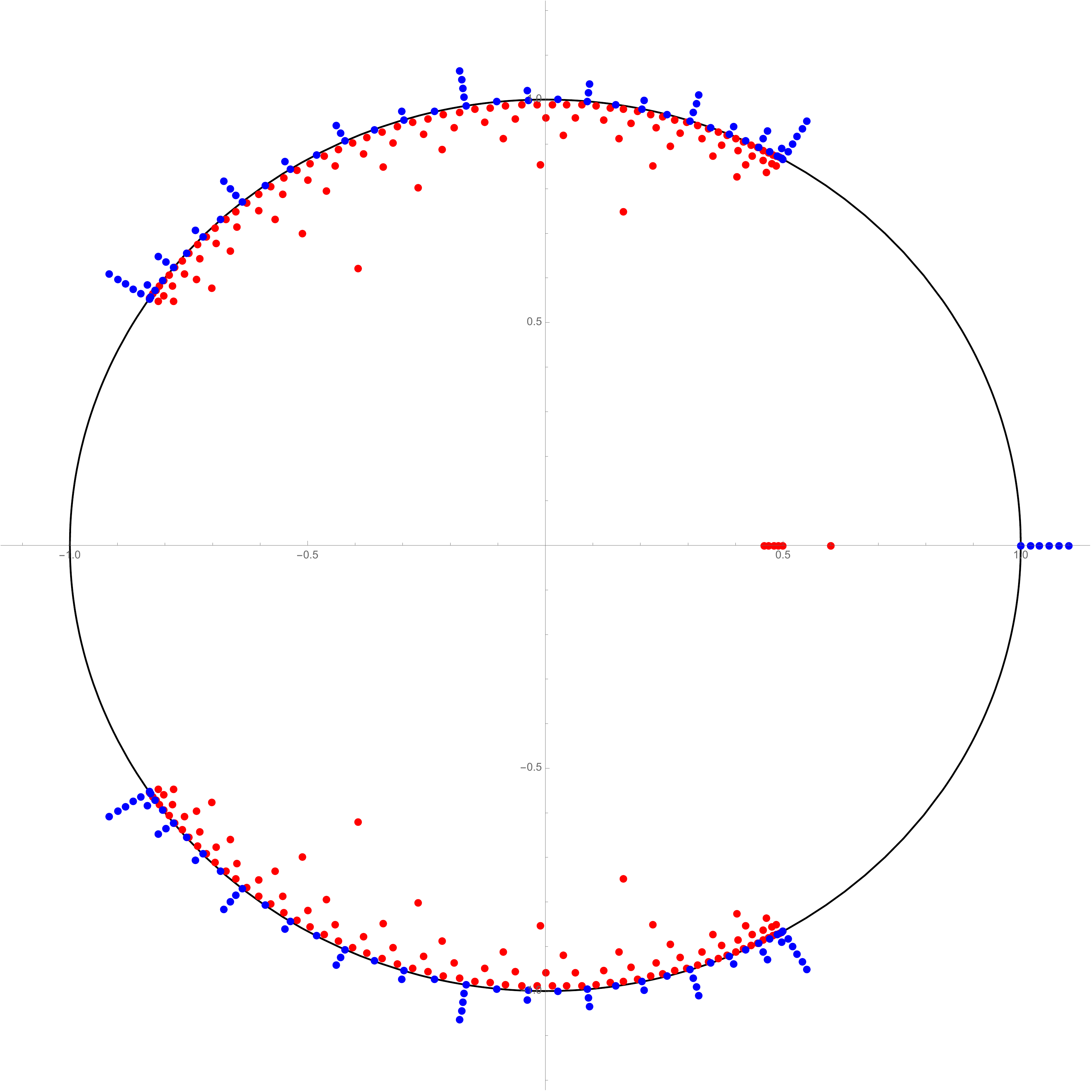}\hfill{}

\hfill{}\includegraphics[bb=0bp 0bp 1000bp 1000bp,clip,width=0.48\columnwidth]{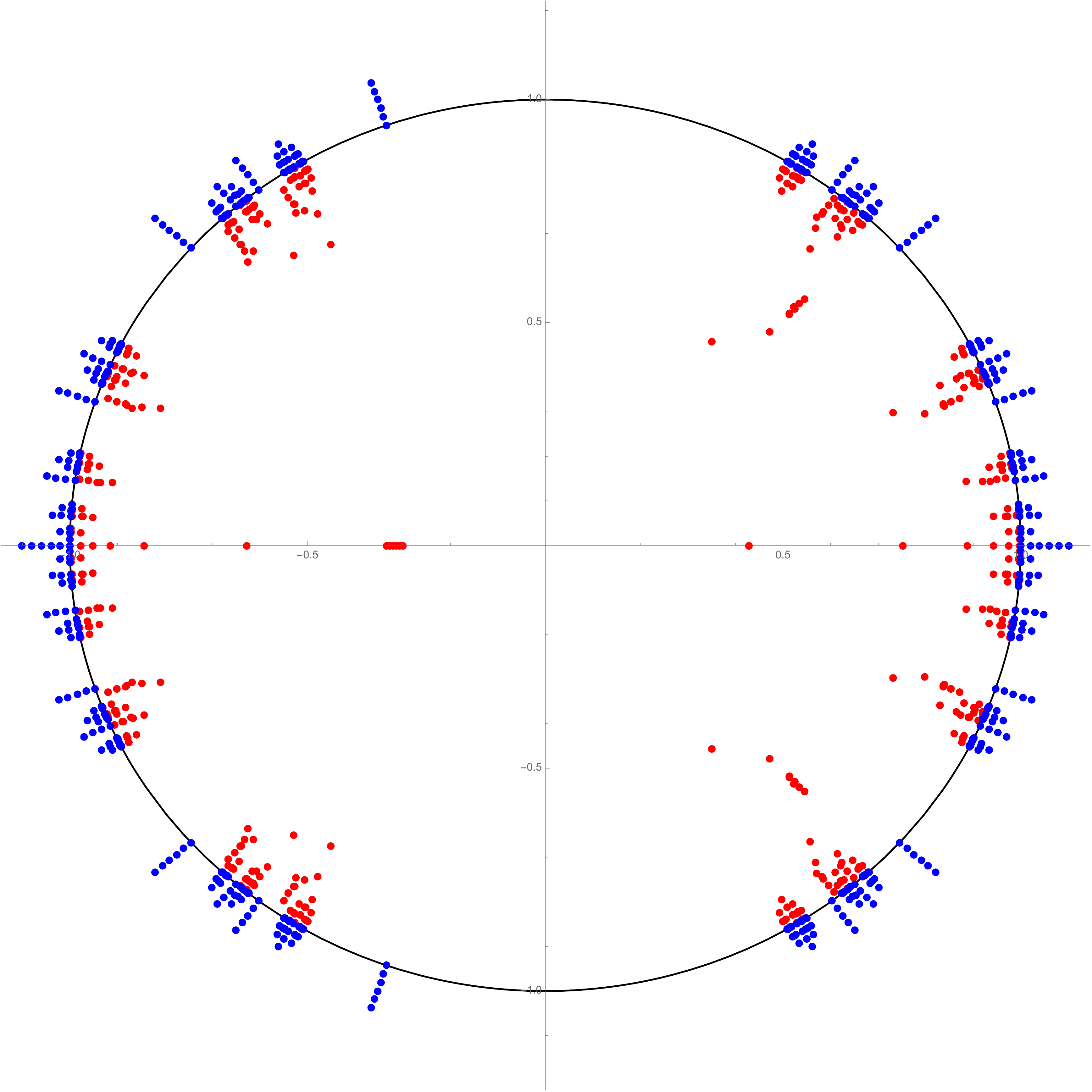}\hfill{}\includegraphics[bb=0bp 0bp 1000bp 1000bp,clip,width=0.48\columnwidth]{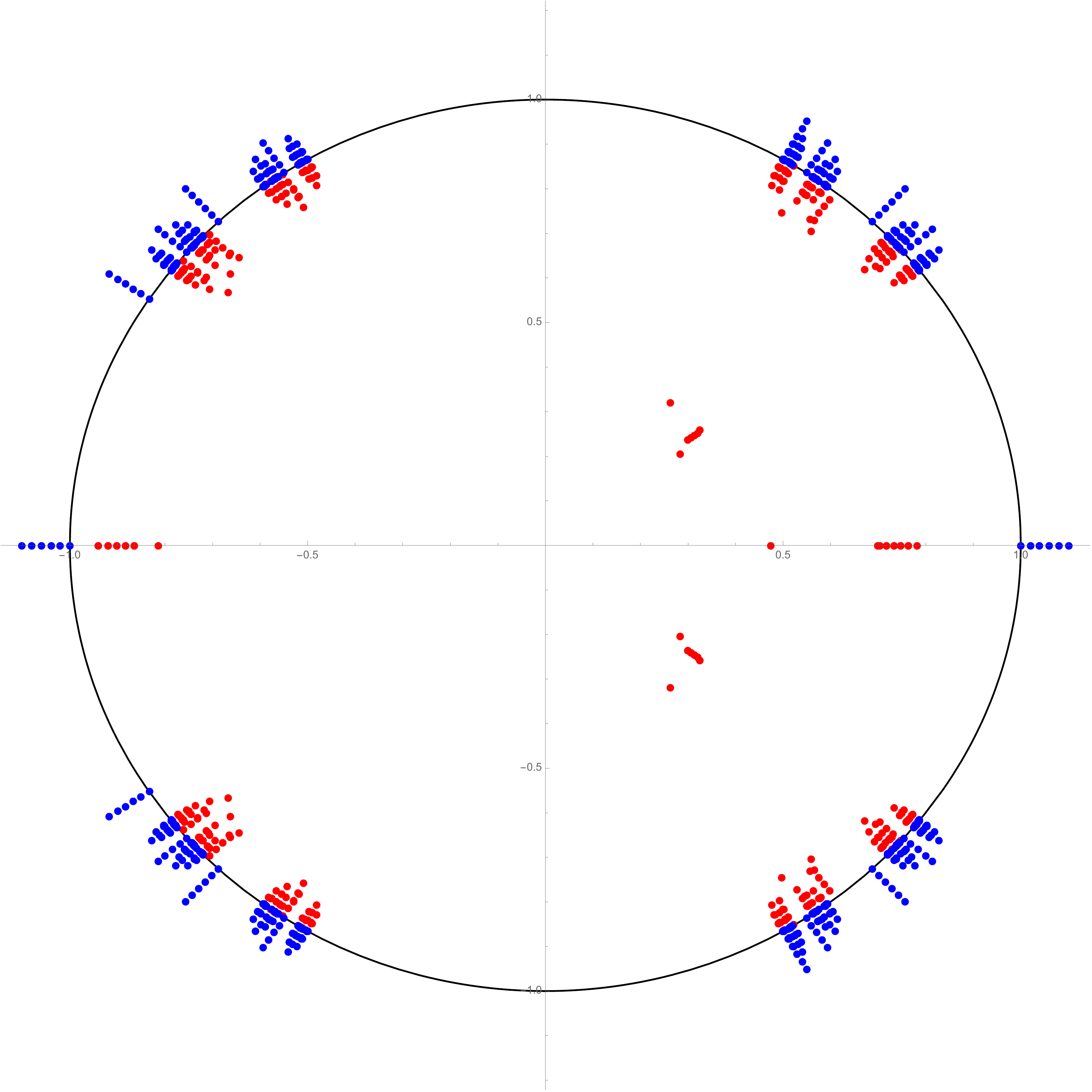}\hfill{}

\caption{\label{fig:DSGpoles}Plot of the poles of the Laplace transforms in
the complex-$z$ plane for the hopping parameters $a_{k}$ and $b_{k}$
(red points), and for the amplitude to remain at the origin, $\overline{\psi}_{0}^{(k)}(z)=X_{k}\psi_{IC}$
(blue points). Shown are the poles for the first few consecutive RG-steps
($k=1,\ldots,6$) for a quantum walk on the 1$d$-line (top) and DSG
(bottom). On the left, the quantum walk is driven by the Grover coin
in Eq.~(\ref{eq:GroverC}), and on the right, the rotational coin
in Eq.~(\ref{eq:nonGroverC}) was applied. Since both coins are real,
poles are certain to occur in complex-conjugate pairs. As $X_{k}$
is unitary, all its poles are on the unit-circle (black line) in the
complex-$z$ plane, however, for better visibility we have offset
poles increasingly outward for smaller $k$. Since the poles of the
hopping parameters are all strictly outside (but converging onto)
the unit-circle, we have applied the same offset but also mapped the
poles (by $z\to1/z$) to the inside. Note that if a pole for the observable
arises at some $k_{0}$, it remains to be a pole for all $k\geq k_{0}$;
this is not true for the hopping parameters. Generally, the pattern
by which poles evolve appears more complicated for DSG. For walks
with the Grover coin ${\cal C}_{G}$, ever more poles progressively
impinge on the real-$z$ axis, at either of $z=\pm1$, or both. For
the non-reflective coin ${\cal C}_{60}$, the accumulation points
for poles generally reside at non-trivial locations , for example
at $\arg z=\pi\pm\arctan\sqrt{35}$ for the 1$d$-line (top-right).
Typically, the edges of the wedges within which poles accumulate represent
fixed points that yield exponents for the classical random walk, such
as at $\arg z=\pi\pm\arctan2\sqrt{2}$ for the 1$d$-line with Grover
coin (top-left), or at both, $\arg z=\pi\pm\arctan\frac{\sqrt{11}}{5}$
and $\arg z=\pm\frac{\pi}{3}$, for the 1$d$-line with the non-reflective
coin (top-right). The domains in which poles appear to become dense
seem to break up like fractal Kantor-sets for the DSG.}
\end{figure*}

Note that the observable $X_{k}$ in Eq.~(\ref{eq:XinChi}) has poles
for $z=1$ and for real $\nu_{n}^{(k)}=n\pi/2^{k-1}$, $0\leq n<2^{k}$.
By Eq.~(\ref{eq:SigNu}), however, $\nu$ can be real only for real
$\omega=\arg z\in I$, i.e., on the corresponding segment of the unit-circle
in the complex-$z$ plane (where we furthermore have a uni-modular
$\xi=e^{i\nu}$, $\left|\xi\right|=1$). The corresponding values
for $\omega_{n}^{(k)}$ can be obtained from Eq.~(\ref{eq:SigNu}).
Those Laplace-poles $\omega_{n}^{(k)}$ are shown in Fig.~\ref{fig:DSGpoles},
obtained directly by iteration of the RG-flow for $k=1,\ldots,6$.
The corresponding poles for the hopping parameters in Eq.~(\ref{eq:akbkinsignu})
are at $\nu_{m}^{(k)}=(m\pi+1/2-\sigma)/2^{k-1}$, $0\leq m<2^{k}$,
pushed slightly off the unit-circle by $\sigma$, as is generally
the case~\cite{Boettcher16}. There is clearly a connection between
those poles and the aforementioned fixed points, as $2^{k}\nu_{n}^{(j)}$
for $k\geq j$ is always an integer multiple of $2\pi$, such that
the corresponding $\omega_{n}^{(j)}$ is also always a fixed point
in Eq.~(\ref{eq:akbkinsignu}). Thus, expanding an observable like
$X_{k}$  for large $k$ (i.e., for any fixed point
$\omega_{n}^{(j)}$ it is assumed $k\gg j$) implies an expansion
around such a pole, and one \emph{generically} finds 
\begin{eqnarray}
X_{k} & \sim\sum_{l=-1}^{\infty} & {\cal M}_{l}\,\left(N\zeta\right)^{l},\qquad\left(\zeta=z-z_{{\rm FP}}\to0\right),\label{eq:Xexpansion}
\end{eqnarray}
where we set $z_{{\rm FP}}=e^{i\omega_{n}^{(j)}}$, and all ${\cal M}_{l}\left(z_{{\rm FP}}\right)$
are matrices that become constant after neglecting higher-order corrections
in inverse size $N=2^{k}$ for each order in $\zeta$. Although those
constants vary with the choice of $z_{{\rm FP}}$, they can not affect
the sought-after scaling behavior. Clearly, $z_{{\rm FP}}=-1$ is
the most convenient (and symmetric, see Fig.~\ref{fig:DSGpoles})
choice. The classical choice $z_{{\rm FP}}=1$ would not work, even
though $z=1$ is a pole of $X_{k}$. However, that pole is isolated,
as Fig.~\ref{fig:DSGpoles} shows, and we observe from Eq.~(\ref{eq:akbkinsignu})
that $\left(a_{k},b_{k}\right)=\left(0,1\right)$ is trivial and without $N$-dependence in its expansion.

Finally, we study how these choices affect the asymptotic
analysis of the RG-flow in Eq.~(\ref{eq:1dRGflowGrover}). After all,
and as we will see below, we are typically not in possession of the
exact location of poles. In some elementary case, like the 2-state
quantum walk on a line~\cite{Boettcher17a}, the RG-flow does not
have an explicit $z$-dependence and the expansion around some $z_{{\rm FP}}$
remains purely formal, without any need for specificity. Even if $z$
explicitly appears, such as in Eq.~(\ref{eq:1dRGflowGrover}), we
can replace $z=z_{{\rm FP}}+\left(z-z_{{\rm FP}}\right)\sim z_{{\rm FP}}$,
since we are aiming to analyze an \emph{unstable} fixed point with
diverging corrections, which corrections in small $\left(z-z_{{\rm FP}}\right)$
can not provide. To wit, for a generic RG-flow $\vec{a}_{k+1}(z)={\cal RG}\left(\vec{a}_{k}(z);z\right)$,
expanded around the fixed point $z_{{\rm FP}}$ in the limit of large
system sizes, $k\to\infty$, we have to linear order:
\begin{eqnarray}
\vec{\alpha}_{k+1}(z) & \sim & \left[\vec{\alpha}_{k}\circ\vec{\nabla}_{\vec{a}}+\left(z-z_{{\rm FP}}\right)\partial_{z}\right]{\cal RG}\left(\vec{a}_{\infty}(z_{{\rm FP}});z_{{\rm FP}}\right)+\ldots,\label{eq:FPlinear}
\end{eqnarray}
with $\vec{\alpha}_{k}(z)=\vec{a}_{k}(z)-\vec{a}_{\infty}(z_{{\rm FP}})$
with $\vec{a}_{\infty}(z_{{\rm FP}})={\cal RG}\left(\vec{a}_{\infty}(z_{{\rm FP}});z_{{\rm FP}}\right)$
defining the fixed point. It is then the divergent eigen-solutions
$\left\{ \lambda_{i},\vec{v}_{i}\right\} $ of the Jacobian $J_{\infty}=\vec{\nabla}_{\vec{a}}{\cal RG}\left(\vec{a}_{\infty}(z_{{\rm FP}});z_{{\rm FP}}\right)$,
i.e., those with eigenvalues $\lambda_{i}>1$, that dominate the solution
for $k\to\infty$, while the $k$-independent inhomogeneity $\left(z-z_{{\rm FP}}\right)\partial_{z}{\cal RG}\left(\vec{a}_{\infty}(z_{{\rm FP}});z_{{\rm FP}}\right)$
remains irrelevant, to this and also in higher orders:
\begin{eqnarray}
\vec{\alpha}_{k}(z) & \sim{\cal A}\lambda_{1}^{k}\vec{v}_{1}+{\cal B}\lambda_{2}^{k}\vec{v}_{2}+\ldots & .\label{eq:Linearization}
\end{eqnarray}
In the case at hand in Eq.~(\ref{eq:1dRGflowGrover}), we find that
the Jacobian is already diagonal such that the eigenvectors are elementary,
$\vec{v}_{i}=\widehat{e}_{i}$, with divergent eigenvectors $\lambda_{1}^{k}=\lambda_{2}^{k}=2^{k}=N$,
and we can insert: 
\begin{eqnarray}
a_{k}\left(z\right) & \sim & a_{\infty}+\zeta{\cal A}\lambda_{1}^{k}+\zeta^{2}\alpha_{k}^{(2)}+\zeta^{3}\alpha_{k}^{(3)}+\ldots,\nonumber \\
b_{k}\left(z\right) & \sim & b_{\infty}+\zeta{\cal B}\lambda_{2}^{k}+\zeta^{2}\beta_{k}^{(2)}+\zeta^{3}\beta_{k}^{(3)}+\ldots,\label{eq:FPansatz}
\end{eqnarray}
while simply setting $z=z_{{\rm FP}}$. We then find as the most-divergent
terms in each order: 
\begin{eqnarray}
\alpha_{k}^{(2)} & \sim & \frac{1+3z_{{\rm FP}}}{2\left(1-z_{{\rm FP}}\right)}\left[\left({\cal A}\lambda_{1}^{k}\right)^{2}+\left({\cal B}\lambda_{2}^{k}\right)^{2}\right]+\ldots,\nonumber \\
\alpha_{k}^{(3)} & \sim & \frac{\left(1+3z_{{\rm FP}}\right)^{2}}{6\left(1-z_{{\rm FP}}\right)^{2}}\left[\left({\cal A}\lambda_{1}^{k}\right)^{3}+5\left({\cal A}\lambda_{1}^{k}\right)\left({\cal B}\lambda_{2}^{k}\right)^{2}\right]+\ldots,\nonumber \\
\beta_{k}^{(2)} & \sim & \frac{1+3z_{{\rm FP}}}{2\left(1-z_{{\rm FP}}\right)}\left({\cal A}\lambda_{1}^{k}\right)\left({\cal B}\lambda_{2}^{k}\right)+\ldots,\label{eq:abterms}\\
\beta_{k}^{(3)} & \sim & \frac{\left(1+3z_{{\rm FP}}\right)^{2}}{3\left(1-z_{{\rm FP}}\right)^{2}}\left[2\left({\cal A}\lambda_{1}^{k}\right)^{2}\left({\cal B}\lambda_{2}^{k}\right)+\left({\cal B}\lambda_{2}^{k}\right)^{3}\right]+\ldots.\nonumber 
\end{eqnarray}
Inserted into $X_{k}$ in Eq.~(\ref{eq:Xkakbk}), we indeed faithfully
reproduce the expansion in Eq.~(\ref{eq:Xexpansion}), up to unknown
(and irrelevant) constants. Not surprisingly, we find that the choice
of $z_{{\rm FP}}$ does not affect the scaling results, except that
$z_{{\rm FP}}=1$ is excluded, consistent with the direct expansion
of the exact solution.

\subsection{RG-Analysis for the Non-reflective Coin\label{subsec:RG-for-NonR}}

We apply the same strategy for the quantum walk with the non-reflective
coin ${\cal C}_{60}$ in Eq.~(\ref{eq:GroverC}): evolving Eqs.\ (\ref{eq:recur1dPRWmass})
for a few iterations from its unrenormalized values~\cite{Boettcher13a},
a recurring pattern emerges that suggests again the Ansatz $A_{k}=a_{k}S^{A}{\cal C}_{60}$,
$B_{k}=a_{k}S^{B}{\cal C}_{60}$, and
\begin{equation}
M_{k}=m_{k}\left[\begin{array}{ccc}
0 & \left(\frac{2-z}{2z-1}\right)b_{k} & 0\\
b_{k} & 0 & 0\\
0 & 0 & z
\end{array}\right]{\cal C}_{60}.\label{eq:MknonGrover}
\end{equation}
Starting from 
\begin{equation}
a_{k=1} = \frac{2z^{2}\left(z-1\right)}{2z-3},\qquad
b_{k=1} = \frac{z^{2}\left(1-2z\right)}{2z-3},\label{eq:RGinitNonG}\\
\end{equation}
the RG-flow now closes after each iteration for 
\begin{eqnarray}
a_{k+1} & = & \frac{2\left(z-1\right)\left(2z-1\right)a_{k}^{2}}{(2z-3)(2z-1)-2(z-2)(2z-1)b_{k}+(z-2)(3z-2)b_{k}^{2}},\label{eq:1dQWRGflowNonG}\\
b_{k+1} & = & b_{k}+\frac{2\left(2z-1\right)a_{k}^{2}\left[\left(3z-2\right)b_{k}-2z+1\right]}{(2z-3)(2z-1)-2(z-2)(2z-1)b_{k}+(z-2)(3z-2)b_{k}^{2}}.\nonumber
\end{eqnarray}
Again, there is a close-form solution for the RG-flow:
\begin{eqnarray}
a_{k} & = & \frac{2\left(z-1\right)\left[\left(z-2\right)\left(3z-2\right)u^{2}-2\left(z-2\right)\left(2z-1\right)u+4z^{2}-8z+3\right]\xi^{2^{k-1}}}{4\left(2z-1\right)\left(z-1\right)^{2}+\left(z-2\right)\left[\left(3z-2\right)u-2z+1\right]^{2}\xi^{2^{k}}},\nonumber \\
b_{k} & = & \frac{2\left(z-1\right)\left\{ 4\left(z-1\right)^{2}u+\left[\left(z-2\right)u-2z+3\right]\left[\left(3z-2\right)u-2z+1\right]\xi^{2^{k}}\right\} }{4\left(2z-1\right)\left(z-1\right)^{2}+\left(z-2\right)\left[\left(3z-2\right)u-2z+1\right]^{2}\xi^{2^{k}}},\label{eq:akbknonG}
\end{eqnarray}
where from Eq.~(\ref{eq:RGinitNonG}) at $k=1$ we get with $z=e^{i\omega}$:
\begin{eqnarray}
\xi & = & \frac{11}{8}+\frac{3}{2}\cos\omega+\frac{9}{4}\cos2\omega-\frac{\sqrt{3}}{8}\left(1+6\cos\omega\right)\sqrt{1+4\cos\omega+6\cos2\omega},\nonumber \\
u & = & e^{i\omega}\frac{2\cos\frac{\omega}{2}-3\cos\frac{3\omega}{2}-i\sqrt{3}\sin\frac{\omega}{2}\sqrt{1+4\cos\omega+6\cos2\omega}}{4\cos\omega-5}.\label{eq:xiUnonG}
\end{eqnarray}
In correspondence with Eq.~(\ref{eq:XiUGrover}), we note that $\xi$
here is complex (with $\left|\xi\right|=1$) for $\omega=\arg z\in I_{\pm}$
on the conjugate intervals $I_{\pm}=\left(\pm\frac{\pi}{3},\pi\mp\arctan\frac{\sqrt{11}}{5}\right)$,
leading to an infinity of solutions for $\omega$ of $\xi^{2^{j}}=1$
for any integer $j\geq0$ form a (likely dense) set of fixed points
of the RG-flow on $I_{\pm}$, because they leave Eq.~(\ref{eq:akbknonG})
invariant for all $j\leq k\leq\infty$. As in the previous case for
a Grover coin, Laplace poles of observables will be found to cluster
on these intervals of the unit-circle in the complex-$z$ plane, see
Fig.~\ref{fig:DSGpoles}. Most remarkable here is the fact that the
real axes, i.e., $z=\pm1$, are not inside those intervals. Extending
this analysis to a new set of variables, $\xi=e^{i\nu}$ and $u=\frac{2\left(z-1\right)}{2-3z}\sqrt{\frac{2z-1}{z-2}}e^{i\sigma}+\frac{2z-1}{3z-2}$,
we find in analogy to Eq.~(\ref{eq:SigNu}):
\begin{equation}
\cos\nu =\frac{11}{8}+\frac{3}{2}\cos\omega+\frac{9}{4}\cos2\omega,\qquad
\sin\sigma = \frac{9z^{3}-3z^{2}-z+2}{4\sqrt{\left(2z-1\right)\left(2-z\right)}}.\label{eq:nusignonG}
\end{equation}
In terms of these variables, we can conveniently express the solutions
in Eq.~(\ref{eq:1dQWRGflowNonG}) as
\begin{eqnarray}
a_{k} & = & \frac{2\left(z-1\right)\cos\sigma}{\left(3z-2\right)\cos\left(2^{k}\nu+\sigma\right)},\label{eq:akbknusignonG}\\
b_{k} & = & \frac{2z-1}{3z-2}+\frac{2\left(z-1\right)\sqrt{2z-1}\sin\left(2^{k}\nu\right)}{\left(3z-2\right)\sqrt{2-z}\cos\left(2^{k}\nu+\sigma\right)}.\nonumber 
\end{eqnarray}

\begin{figure}

\hfill{}\includegraphics[bb=0bp 180bp 792bp 612bp,clip,width=0.8\columnwidth]{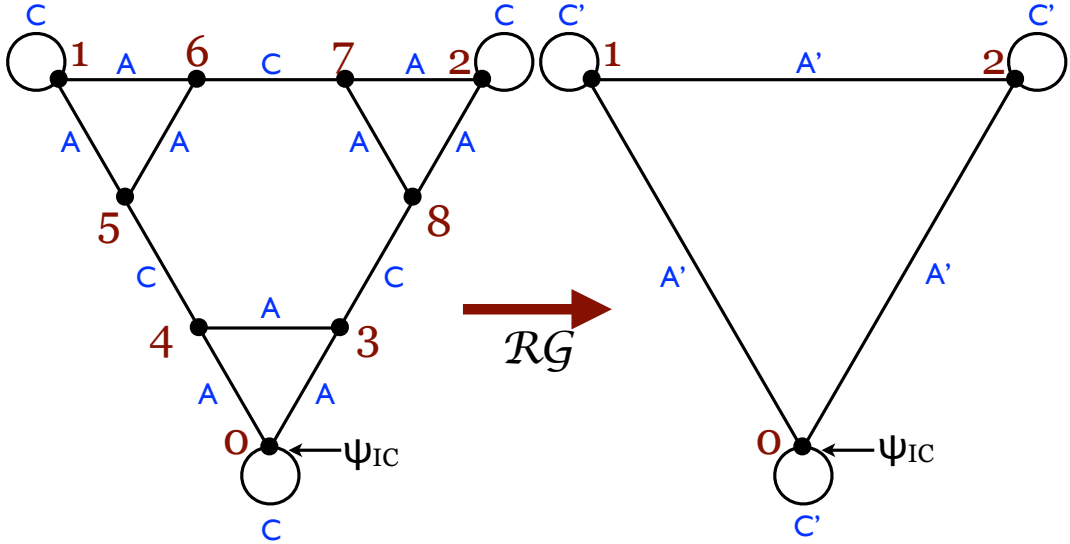}\hfill{}

\caption{\label{fig:DSGRGstep}Depiction of the (final) RG-step in the analysis
of DSG. The letters $\left\{ A,C\right\} $ label transitions between
sites (black dots on the vertices) of the quantum walk in the form
of hopping matrices. (Only on the three outermost sites, the matrix
$C$ refers back to the same site. For simplicity, we have omitted
all such self-loops associated with hopping matrix $M$ at each site.)
Recursively, the inner-6 sites (labeled $3,\ldots,8$) of each larger
triangle (left) in DSG are decimated to obtain a reduced triangle
(right) with renormalized hopping matrices (primed). To build a DSG
of $N=3^{g}$ sites, this procedure is applied (in reverse) $g$ times
to all triangles at each generation. Note that each generation the
base-length $L$ increases by a factor of $b=2$, such that the fractal
dimension is $d_{f}=\log_{b=2}3$.}
\end{figure}

Using Eq.~(\ref{eq:XkDef}), we obtain the observable $X_{k}$, which
in terms of Eq.~(\ref{eq:akbknusignonG}) takes on the same form as
in Eq.~(\ref{eq:XinChi}) (with an equivalent but numerically different
matrix $\chi$). All of $a_{k}$, $b_{k}$, and $X_{k}$ appear very
similar to their counterparts derived from the Grover coin in 
Sec.~\ref{subsec:RG-for-Grover}, with poles of $X_{k}$ for $z=1$ and
for real $\nu_{n}^{(k)}=n\pi/2^{k-1}$, $0\leq n<2^{k}$. In terms
of the location of the poles in the complex-$z$ plane these values
are attained via $\cos\nu$ as given in Eqs.~(\ref{eq:SigNu}) and~(\ref{eq:nusignonG}), 
respectively. Accordingly, the $\nu_{n}^{(k)}$
get mapped onto the unit-circle with $\omega=\arg z$ in the intervals
$I$ and $I_{\pm}$, respectively, as shown in Fig.~\ref{fig:DSGpoles},
knowledge of which would be essential to expand the exact $X_{k}$
around some pole within the respective intervals. In contrast, the
expansion of the RG-flow around some putative fixed point $z_{{\rm FP}}$,
for either the non-reflective coin just or for the Grover coin in Eq.~(\ref{eq:FPansatz}), 
is \emph{oblivious} to the choice of $z_{{\rm FP}}$
in so far as the sought-after scaling behavior is concerned {[}i.e.,
aside from any numerical factors dependent on $z_{{\rm FP}}$, as
in Eq.~(\ref{eq:abterms}){]}. However,  exceptional points may
exist, such as the isolated pole at $z=1$ in both examples, that
 should be avoided.

Despite the  similarities in the analysis in both cases, the
shift in poles away from the real-$z$ axis could have signaled a
change in universality classes for the quantum walk with these very
distinct coins. Yet, in this simple example of a \emph{1d}-walk, we
find that both RG-flows provide identical eigenvalues for their Jacobians,
leaving the walk dimension $d_{w}^{Q}$ in Eq.~(\ref{eq:dwQ}) unchanged.
In the following, we will apply these insights to the more involved
(and not exactly solvable) RG-flow for a quantum walk on a dual Sierpinski
gasket (DSG). 

\section{Quantum Walk on the Dual Sierpinski Gasket\label{sec:Quantum-Walk-DSG}}

Basically, the same procedure as in Sec.~\ref{sec:RG-for-the},
although algebraically more laborious, can be applied to obtain closed
RG-recursions for a number of fractal networks~\cite{QWNComms13,Boettcher14b}.
These have provided the first exact weak-limit results~\cite{grimmett_2004a,Konno05,Falkner14a}
for non-trivial networks beyond regular lattices. 

Due to the self-similarity of fractal networks, we can decompose ${\cal U}_{x,y}$
in Eq.~(\ref{eq:z_master}) into its smallest sub-structure, exemplified
by Fig.~\ref{fig:DSGRGstep}. It shows the elementary graph-let of
nine sites that is used to recursively build the dual Sierpinski gasket
(DSG) of size $N=3^{g}$ after $g$ generations. The master equations
pertaining to these sites are: 
\begin{eqnarray}
\overline{\psi}_{0} & = & \left(M+C\right)\overline{\psi}_{0}+A\left(\overline{\psi}_{3}+\overline{\psi}_{4}\right)+\psi_{IC},\nonumber \\
\overline{\psi}_{\left\{ 1,2\right\} } & = & \left(M+C\right)\overline{\psi}_{\left\{ 1,2\right\} }+A\left(\overline{\psi}_{\left\{ 5,7\right\} }+\overline{\psi}_{\left\{ 6,8\right\} }\right),\label{eq:DSG_master}\\
\overline{\psi}_{\left\{ 3,4,5,6,7,8\right\} } & = & M\overline{\psi}_{\left\{ 3,4,5,6,7,8\right\} }+C\overline{\psi}_{\left\{ 8,5,4,7,6,3\right\} }+A\left(\overline{\psi}_{\left\{ 0,3,1,5,2,7\right\} }+\overline{\psi}_{\left\{ 4,0,6,1,8,2\right\} }\right).\nonumber 
\end{eqnarray}
The hopping matrices $A$ and $C$ describe transitions between neighboring
sites, while $M$ permits the walker to remain on its site in a ``lazy''
walk. The inhomogeneous $\psi_{IC}$-term allows for an initial condition
$\psi_{x,t=0}=\delta_{x,0}\psi_{IC}$ for a quantum walker to start
at site $x=0$ in state $\psi_{IC}$. 

\subsection{Renormalization Group\label{sec:Renormalization-Group-(RG)}}

To accomplish the decimation of the sites $\overline{\psi}_{\left\{ 3,\ldots,8\right\} }$,
as indicated in Fig.~\ref{fig:DSGRGstep}, we need to solve the linear
system in Eqs.~(\ref{eq:DSG_master}) for $\overline{\psi}_{\left\{ 0,1,2\right\} }$.
Thus, we expect that $\overline{\psi}_{\left\{ 3,\ldots,8\right\} }$
can be expressed as (symmetrized) linear combinations 
\begin{eqnarray}
\overline{\psi}_{\left\{ 3,4\right\} } & = & P\overline{\psi}_{0}+Q\overline{\psi}_{\left\{ 1,2\right\} }+R\overline{\psi}_{\left\{ 2,1\right\} },\nonumber \\
\overline{\psi}_{\left\{ 5,8\right\} } & = & R\overline{\psi}_{0}+P\overline{\psi}_{\left\{ 1,2\right\} }+Q\overline{\psi}_{\left\{ 2,1\right\} },\label{eq:Ansatz}\\
\overline{\psi}_{\left\{ 6,7\right\} } & = & Q\overline{\psi}_{0}+P\overline{\psi}_{\left\{ 1,2\right\} }+R\overline{\psi}_{\left\{ 2,1\right\} }.\nonumber 
\end{eqnarray}
Inserting this Ansatz into Eqs.~(\ref{eq:DSG_master}) and comparing
coefficients provides consistently for the unknown matrices: 
\begin{eqnarray}
P & = & \left(M+A\right)P+A+CR,\nonumber \\
Q & = & \left(M+C\right)Q+AR,\label{eq:PQRJ}\\
R & = & MR+AQ+CP.\nonumber 
\end{eqnarray}
Using the abbreviations $S=\left(\mathbb{I}-M-C\right)^{-1}A$ and
$T=\left(\mathbb{I}-M-AS\right)^{-1}C$, Eqs.~(\ref{eq:PQRJ}) have
the solution: 
\begin{equation}
P=\left(\mathbb{I}-M-A-CT\right)^{-1}A,\qquad R=TP,\qquad Q=SR.\label{eq:solPQRJ}
\end{equation}

Finally, after $\overline{\psi}_{\left\{ 3,\ldots,8\right\} }$ have
been eliminated, we find 
\begin{align}
\overline{\psi}_{0} & =\left(\left[M+2AP\right]+C\right)\overline{\psi}_{0}+A\left(Q+R\right)\left(\overline{\psi}_{1}+\overline{\psi}_{2}\right)+\psi_{IC},\label{eq:psi0RG}\\
\overline{\psi}_{\left\{ 1,2\right\} } & =\left(\left[M+2AP\right]+C\right)\overline{\psi}_{\left\{ 1,2\right\} }+A\left(Q+R\right)\left(\overline{\psi}_{0}+\overline{\psi}_{\left\{ 2,1\right\} }\right).\nonumber
\end{align}
By comparing coefficients between the renormalized expression in Eq.~(\ref{eq:psi0RG})
and the corresponding, \emph{self-similar} expression in the first
line of Eqs.~(\ref{eq:DSG_master}), we can identify the RG-recursions
\begin{eqnarray}
M_{k+1} & = & M_{k}+2A_{k}P_{k},\nonumber \\
A_{k+1} & = & A_{k}\left(Q_{k}+R_{k}\right),\nonumber \\
C_{k+1} & = & C_{k},\label{eq:RGrecur}
\end{eqnarray}
where the subscripts refer to $k$- and $(k+1)$-renormalized forms
of the hopping matrices. These recursions evolve from the un-renormalized
($k=0$) hopping matrices with $\left\{ M,A,C\right\} _{k=0}  =  z\left\{ M,A,C\right\}$. 
These RG-recursions are entirely generic and, in fact, would hold
for any walk on DSG, classical or quantum. In the following, we now
consider a specific form of a quantum walk with both, the Grover coin
in Eq.~(\ref{eq:GroverC}) as well as the non-reflective coin in Eq.
(\ref{eq:nonGroverC}).

As an observable, we focus again on $\overline{\psi}_{0}(z)$, the
amplitude at the origin of the quantum walk on DSG, here chosen on
one of the corner-sites. According to Fig.~\ref{fig:DSGRGstep} and
Eq.~(\ref{eq:psi0RG}), we have 
\begin{eqnarray}
\overline{\psi}_{0} & = & \left(M_{k}+C_{k}\right)\overline{\psi}_{0}+A_{k}\left(\overline{\psi}_{1}+\overline{\psi}_{2}\right)+\psi_{IC},\label{eq:DSGfinal}\\
\overline{\psi}_{\left\{ 1,2\right\} } & = & \left(M_{k}+C_{k}\right)\overline{\psi}_{\left\{ 1,2\right\} }+A_{k}\left(\overline{\psi}_{0}+\overline{\psi}_{\left\{ 2,1\right\} }\right),\nonumber 
\end{eqnarray}
which has the solution $\overline{\psi}_{0}=X_{k}\psi_{IC}$ with
\begin{equation}
X_{k}=\left[\mathbb{I}-M_{k}-C_{k}-2A_{k}\left(\mathbb{I}-M_{k}-A_{k}-C_{k}\right)^{-1}A_{k}\right]^{-1}.\label{eq:Xmatrix}
\end{equation}
Preserving the norm of the quantum walk demands unitary propagation,
i.e., $\mathbb{I}={\cal U}^{\dagger}{\cal U}$. In Ref.~\cite{Boettcher17b},
we have obtained generalized unitarity conditions for DSG as shown
in Fig.~\ref{fig:DSGRGstep}:
\begin{eqnarray}
\mathbb{I} & = & 2A^{\dagger}A+C^{\dagger}C+M^{\dagger}M,\label{eq:UnitarityDSG}\\
0 & = & A^{\dagger}A+A^{\dagger}M+M^{\dagger}A=C^{\dagger}M+M^{\dagger}C=A^{\dagger}C.\nonumber 
\end{eqnarray}

\subsection{RG-Analysis for the Quantum Walk with Grover Coin \label{sec:The-Quantum-Walk-DSG}}

In this section, we review the behavior of a coined quantum walk on
the DSG with the Grover-coin ${\cal C}_{G}$ in Eq.~(\ref{eq:GroverC}),
adapting the analysis from Ref.~\cite{Boettcher17a}. To study the
scaling solution for the spreading quantum walk according to Eq.~(\ref{eq:collapse}),
as expressed by the walk dimension in Eq.~(\ref{eq:dwQ}), it is sufficient
to investigate the properties of the RG-recursion in the previous
section for $\left\{ M,A,C\right\} $. In the unrenormalized (``raw'')
description of the walk, these hopping matrices are chosen as 
\begin{align}
M=\left[\begin{array}{ccc}
-\frac{1}{3} & 0 & 0\\
0 & -\frac{1}{3} & 0\\
0 & 0 & 0
\end{array}\right]{\cal C}_{G}, & \quad A=\left[\begin{array}{ccc}
\frac{2}{3} & 0 & 0\\
0 & \frac{2}{3} & 0\\
0 & 0 & 0
\end{array}\right]{\cal C}_{G},\quad C=\left[\begin{array}{ccc}
0 & 0 & 0\\
0 & 0 & 0\\
0 & 0 & 1
\end{array}\right]{\cal C}_{G}.\label{eq:MAC_IC}
\end{align}
Here, we have to pay a small price for the fact that throughout, $A$
shifts weights \emph{symmetrically} to two neighboring sites within
their local triangle. To maintain the unitarity conditions in Eq.
(\ref{eq:UnitarityDSG}), the walk now must have a ``lazy'' component,
i.e., some weight may remain at each site every update, so that $M\not=0$.
The matrix $C$ shifts weight to the one neighbor outside those triangles,
as illustrated in Fig.~\ref{fig:DSGRGstep}. These weights are the
three complex components of the state vector at each site, $\psi_{x,t}$,
which are all zero at $t=0$, except for $x=0$ where $\psi_{0,0}=\psi_{IC}$
is arbitrary but normalized, $\left|\psi_{IC}^{2}\right|=1$. These
weights are mixed via ${\cal C}_{G}$ at each site before
a shift entangles them within the network.

Iterating the RG-recursions in Eq.~(\ref{eq:RGrecur}) for the matrices
in Eq.~(\ref{eq:MAC_IC}) for only $k=2$ steps already reveals a
simple recursive pattern that suggests the Ansatz 
\begin{align}
M_{k}=\left[\begin{array}{ccc}
\frac{a_{k}-2b_{k}}{3} & \frac{a_{k}-2b_{k}}{3}+\frac{3z+1}{3/z+1} & 0\\
\frac{a_{k}-2b_{k}}{3}+\frac{3z+1}{3/z+1} & \frac{a_{k}-2b_{k}}{3} & 0\\
0 & 0 & 0
\end{array}\right]{\cal C}_{G},\quad & A_{k}=\left[\begin{array}{ccc}
\frac{a_{k}+b_{k}}{3} & \frac{a_{k}+b_{k}}{3} & 0\\
\frac{a_{k}+b_{k}}{3} & \frac{a_{k}+b_{k}}{3} & 0\\
0 & 0 & 0
\end{array}\right]{\cal C}_{G},\label{eq:MAC_k}
\end{align}
while $C_{k}=C_{0}=zC$ does not renormalize. Inserted into the RG-recursions
in Sec.~\ref{sec:Renormalization-Group-(RG)}, these matrices exactly
reproduce themselves \emph{in form} after one iteration, $k\to k+1$,
when we identify for the scalar RG-flow:
\begin{eqnarray}
a_{k+1} & = & \frac{\left(9z^{3}+5z^{2}-5z-9\right)\left(a_{k}-2b_{k}\right)-6\left(3z^{2}+10z+3\right)a_{k}b_{k}}{3\left(9z^{3}+5z^{2}-5z-9\right)+2\left(3z^{2}+10z+3\right)\left(2a_{k}-b_{k}\right)},\nonumber \\
\label{eq:RGflowDSG}\\
b_{k+1} & = & \frac{\begin{array}{l}
4(3+z)\left(9-4z-14z^{2}+6z^{3}+27z^{4}\right)\left(2a_{k}-b_{k}\right)b_{k}-12(3+z)^{2}\left(9z^{2}-1\right)a_{k}b_{k}^{2}\\
\quad-\left(z^{2}-1\right)\left(9-10z+9z^{2}\right)\left(9+14z+9z^{2}\right)\left(a_{k}-2b_{k}\right)-24z(3+z)^{2}b_{k}^{2}
\end{array}}{\begin{array}{l}
4(3+z)\left(9-10z-16z^{2}+6z^{3}+27z^{4}\right)a_{k}+3\left(z^{2}-1\right)\left(9-10z+9z^{2}\right)\left(9+14z+9z^{2}\right)\\
\quad-8(3+z)\left(9+8z-10z^{2}+6z^{3}+27z^{4}\right)b_{k}-4(3+z)^{2}\left(9z^{2}-1\right)\left(2a_{k}-b_{k}\right)b_{k}
\end{array}}.\nonumber 
\end{eqnarray}
 This flow is initiated at $k=1$ with 
\begin{equation}
a_{k=1} = \frac{z(z-1)\left(9+22z+9z^{2}\right)}{(3+z)\left(9+2z-3z^{2}\right)},\qquad b_{k=1} = \frac{z(1+z)\left(9+4z+6z^{2}+4z^{3}+9z^{4}\right)}{(3+z)\left(9+2z+4z^{2}-2z^{3}+3z^{4}\right)}.
\label{eq:abIC}\\
\end{equation}

Inserting Eq.~(\ref{eq:MAC_k}) into Eq.~(\ref{eq:Xmatrix}), we then
also have an expression for the observable $X_{k}$ in terms of the
hopping parameters $a_{k}$ and $b_{k}$. Here, of course, we don't
have any closed-form solution of the RG-flow in Eq.~(\ref{eq:RGflowDSG}).
We can merely verify numerically with a few iterations of the RG-flow
that the Laplace-poles of the hopping parameters as well as of the
observable behave quite similarly as those for the Grover-coin walk
on the \emph{1d}-line, see Fig.~\ref{fig:DSGpoles}. Thus, in the
case of the DSG, we only have the RG to rely on.

\subsubsection{Fixed-Point Analysis\label{sec:RG-Analysis-for-the}}

We now proceed to study the fixed-point properties of the RG-flow
at $k\sim k+1\to\infty$. Eq.~(\ref{eq:RGflowDSG}) has a non-trivial
fixed point for $a_{\infty}\left(z\right)=\frac{\left(1-z\right)\left(9+14z+9z^{2}\right)}{2\left(z+3\right)\left(3z-1\right)}$
and $b_{\infty}(z)=\frac{\left(1+z\right)\left(9-10z+9z^{2}\right)}{2\left(z+3\right)\left(3z-1\right)}$.
Due to the choice of $a_{k}$ and $b_{k}$ in Eq.~(\ref{eq:MAC_k}),
the Jacobian matrix $J_{\infty}=\left.\frac{\partial\left(a_{k+1},b_{k+1}\right)}{\partial\left(a_{k},b_{k}\right)}\right|_{k\to\infty}$
of the fixed point is already diagonal and $z$-independent, with
two eigenvalues, $\lambda_{1}=3$ and $\lambda_{2}=\frac{5}{3}$,
reproducing via Eq.~(\ref{eq:dwQ}) the already known result for the
quantum walk dimension of the DSG with a Grover coin, $d_{w}^{Q}=\log_{2}\sqrt{5}$
~\cite{Boettcher14b}. Extending the expansion of Eq.~(\ref{eq:RGflowDSG})
in powers of $\zeta=z-z_{{\rm FP}}$ for some unspecified fixed-point
value $z_{{\rm FP}}$ for $k\to\infty$ to higher order, we obtain:
\begin{eqnarray}
a_{k}\left(z\right) & \sim & a_{\infty}\left(z_{{\rm FP}}\right)+\zeta^{1}{\cal A}\lambda_{1}^{k}+\zeta^{2}\alpha_{k}^{(2)}+\zeta^{3}\alpha_{k}^{(3)}+\ldots,\nonumber \\
b_{k}\left(z\right) & \sim & b_{\infty}\left(z_{{\rm FP}}\right)+\zeta^{1}{\cal B}\lambda_{2}^{k}+\ldots,\label{eq:abFP}
\end{eqnarray}
with unknown constants ${\cal A}$ and ${\cal B}$, and with 
\begin{eqnarray}
\alpha_{k}^{(2)} & \sim & c\left({\cal A}\lambda_{1}^{k}\right)^{2}+\ldots,\label{eq:alphak}\\
\alpha_{k}^{(3)} & \sim & c^{2}\left[\left({\cal A}\lambda_{1}^{k}\right)^{3}-\frac{1}{2}\left({\cal A}\lambda_{1}^{k}\right)^{2}\left({\cal B}\lambda_{2}^{k}\right)\right]+\ldots,\nonumber 
\end{eqnarray}
with $c=\left(9z_{{\rm FP}}^{2}-1\right)/\left(8z_{{\rm FP}}\right)$,
where we have only kept leading-order terms in $k$ that contribute
to leading order in large $k$. Note that $z_{{\rm FP}}$ has no effect
on the scaling and merely contributes numerically for any choice of
$z_{{\rm FP}}$ on the unit-circle, irrespective of the location of
any poles. Inserting Eqs.~(\ref{eq:MAC_k}) and (\ref{eq:abFP}) into $X_{k}$
in Eq.~(\ref{eq:Xmatrix}) and expanding (some generic component of
$X_{k}$) in powers of $\zeta=z-z_{{\rm FP}}$ yields 
\begin{eqnarray}
\left[X_{k}\right]_{11} & \sim & -\zeta^{-1}\frac{z_{{\rm FP}}+3}{12\left(3z_{{\rm FP}}+1\right)\left({\cal A}\lambda_{1}^{k}\right)}-\zeta^{0}\left[\frac{\left(z_{{\rm FP}}+3\right)\left(3-25z_{{\rm FP}}-3z_{{\rm FP}}^{2}+9z_{{\rm FP}}^{3}\right)}{96z_{{\rm FP}}\left(z_{{\rm FP}}^{2}-1\right)}\right]\label{eq:X11zeta}\\
 &  & -\zeta^{1}\left[\frac{5\left(z_{{\rm FP}}+3\right)\left(3z_{{\rm FP}}-1\right)^{2}\left(3z_{{\rm FP}}+1\right)}{1536z_{{\rm FP}}^{2}}\left({\cal B}\lambda_{2}^{k}\right)\right]+\ldots,\nonumber \\
 & \sim & \zeta^{-1}O\left(\frac{1}{\lambda_{1}^{k}}\right)+\zeta^{0}O(1)+\zeta^{1}O\left(\lambda_{2}^{k}\right)+\ldots.\nonumber 
\end{eqnarray}
While this result reproduces those from Ref.~\cite{Boettcher17a},
it is remarkable that in this case we have to exclude the values $z_{{\rm FP}}=\pm1$,
although those seem to be squarely inside the domain of poles, see
Fig.~\ref{fig:DSGpoles}.

\subsection{RG-Analysis for the Quantum Walk with a Non-Reflective Coin \label{sec:The-Quantum-Walk-DSG-1}}

Here, we follow the script from the previous section, except that
the quantum walk in this case is driven by the non-reflective coin
${\cal C}_{60}$ in Eq.~(\ref{eq:nonGroverC}). In this way, we can
scrutinize some of the analytic features we found in prior sections.
However, it is also the first attempt to assess the impact the reflectivity
of coins has on the evolution of a quantum walk in a non-trivial geometry.
Does breaking of symmetry in such an internal degree of freedom affect
the universality, as would be expressed in a change of the walk dimension
$d_{w}^{Q}$ in Eq.~(\ref{eq:collapse}), for example? 

Again, we investigate the properties of the RG-recursion of Sec.~\ref{sec:Quantum-Walk-DSG}
for $\left\{ M,A,C\right\} $. In the unrenormalized (``raw'') description
of the walk, these hopping matrices are 
\begin{align}
M=\left[\begin{array}{ccc}
-\frac{1}{3} & 0 & 0\\
0 & -\frac{1}{3} & 0\\
0 & 0 & 0
\end{array}\right]{\cal C}_{60}, & \quad A=\left[\begin{array}{ccc}
\frac{2}{3} & 0 & 0\\
0 & \frac{2}{3} & 0\\
0 & 0 & 0
\end{array}\right]{\cal C}_{60},\quad C=\left[\begin{array}{ccc}
0 & 0 & 0\\
0 & 0 & 0\\
0 & 0 & 1
\end{array}\right]{\cal C}_{60},\label{eq:MAC_ICnonG}
\end{align}
the same matrices as in Eq.~(\ref{eq:MAC_IC}), except for the change
of coin. We need to iterate the RG-recursions in Eq.~(\ref{eq:RGrecur})
for the matrices in Eq.~(\ref{eq:MAC_ICnonG}) for only $k=2$ steps
to find a recursive pattern:
\begin{align}
M_{k} & =\left(\frac{a_{k}-2b_{k}}{3}\left[\begin{array}{ccc}
1 & \frac{3-2z}{z-6} & 0\\
\frac{z-6}{3-2z} & 1 & 0\\
0 & 0 & 0
\end{array}\right]+\left[\begin{array}{ccc}
0 & \frac{z\left(1-6z\right)}{z-6} & 0\\
\frac{z\left(2-3z\right)}{2z-3} & 0 & 0\\
0 & 0 & 0
\end{array}\right]\right){\cal C}_{60},\label{eq:MAC_knonG}\qquad
A_{k} & =\frac{a_{k}+b_{k}}{3}\left[\begin{array}{ccc}
1 & \frac{3-2z}{z-6} & 0\\
\frac{z-6}{3-2z} & 1 & 0\\
0 & 0 & 0
\end{array}\right]{\cal C}_{60},
\end{align}
and $C_{k}=C_{0}=zC$, as before. Inserted into the RG-recursions
in Sec.~\ref{sec:Renormalization-Group-(RG)}, these matrices again
exactly reproduce themselves when we identify for the scalar RG-flow:
\begin{eqnarray}
a_{k+1} & = & \frac{2\left(9z^{2}-z+9\right)\left(z^{2}-1\right)\left(a_{k}-2b_{k}\right)-3\left(12z^{3}-41z^{2}-23z+3\right)a_{k}b_{k}}{6\left(9z^{2}-z+9\right)\left(z^{2}-1\right)+\left(12z^{3}-41z^{2}-23z+3\right)\left(2a_{k}-b_{k}\right)},
\label{eq:RGflowDSG_C60}\\
b_{k+1} & = & \frac{\begin{array}{l}
3\left(3+7z-49z^{2}+12z^{3}\right)\left(3-23z-41z^{2}+12z^{3}\right)a_{k}b_{k}^{2}+4\left(z^{2}-1\right)^{2}\left(9-7z+9z^{2}\right)\left(9-z+9z^{2}\right)\left(a_{k}-2b_{k}\right)\\
\quad+4\left(z^{2}-1\right)\left(27-48z-331z^{2}+208z^{3}-453z^{4}+108z^{5}\right)\left(b_{k}-2a_{k}\right)b_{k}-24z\left(z^{2}-1\right)\left(2z-3\right)\left(z-6\right)b_{k}^{2}
\end{array}}{\begin{array}{l}
\left(3-23z-41z^{2}+12z^{3}\right)\left(3+7z-49z^{2}+12z^{3}\right)\left(2a_{k}-b_{k}\right)b_{k}-12\left(z^{2}-1\right)^{2}\left(9-7z+9z^{2}\right)\left(9-z+9z^{2}\right)\\
\quad-4\left(z^{2}-1\right)\left(27-12z-361z^{2}+212z^{3}-453z^{4}+108z^{5}\right)\left(a_{k}-2b_{k}\right)-48z\left(z^{2}-1\right)\left(2z-3\right)\left(z-6\right)b_{k}
\end{array}},\nonumber 
\end{eqnarray}
now initialized by
\begin{equation}
a_{k=1} = \frac{z(z-1)\left(9-17z+9z^{2}\right)}{27-30z+14z^{2}-3z^{3}},\qquad b_{k=1} =\frac{z(z^{2}-1)\left(9-8z-15z^{2}-8z^{3}+9z^{4}\right)}{-27-24z+16z^{2}+28z^{3}-8z^{5}+3z^{6}}.
\label{eq:abIC_C60}
\end{equation}

For the observable $X_{k}$, defined in Eq.~(\ref{eq:Xmatrix}), we
iterate the RG-recursions in Sec.~\ref{sec:Renormalization-Group-(RG)}
symbolically as a function of $z$ up to $k=6$ and numerically determine
the poles of the hopping parameters (all of which have a common denominator)
and of $X_{k}$. These are again plotted in Fig.~\ref{fig:DSGpoles}.
These Laplace-poles share the clustering in fractalized domains already
observed for the DSG with the Grover coin, however, these domains
do not include the real-$z$ axis, similar to the case of using the
same non-reflective coin ${\cal C}_{60}$ on a \emph{1d}-line. Drawing
on both of these cases as reference, we need not worry about a specific
choice of a fixed point $z_{{\rm FP}}$ and whether it is close to
$z=\pm1$. A generic fixed-point analysis should suffice to uniquely
determine the universality class of this quantum walk. 

\subsubsection{Fixed-Point Analysis\label{sec:RG-Analysis-for-the-2}}

Here, the fixed-point properties of the RG-flow in Eq.~(\ref{eq:RGflowDSG})
has a non-trivial fixed point for $a_{\infty}\left(z\right)=\frac{2\left(1-z^{2}\right)\left(9-z+9z^{2}\right)}{3-23z-41z^{2}+12z^{3}}$
and $b_{\infty}(z)=\frac{2\left(z^{2}-1\right)\left(9-7z+9z^{2}\right)}{3+7z-49z^{2}+12z^{3}}$.
As before, $a_{k}$ and $b_{k}$ in Eq.~(\ref{eq:MAC_k}) were chosen
so that the Jacobian matrix $J_{\infty}$ is diagonal, with the same
two eigenvalues, $\lambda_{1}=3$ and $\lambda_{2}=\frac{5}{3}$,
obtained for the Grover coin. Thus, breaking the coin's reflectivity
appears to leave the universality class of this quantum walk unaffected,
as expressed by $d_{w}^{Q}$. 

Extending the expansion of Eq.~(\ref{eq:RGflowDSG}) in powers of
$\zeta=z-z_{{\rm FP}}$ for some unspecified fixed-point value $z_{{\rm FP}}$
for $k\to\infty$ to higher order, we obtain again Eqs.~(\ref{eq:abFP}-\ref{eq:alphak})
but with pre-factor
\begin{eqnarray}
c & = & \frac{\left(3+7z_{{\rm FP}}-49z_{{\rm FP}}^{2}+12z_{{\rm FP}}^{3}\right)\left(3-23z_{{\rm FP}}-41z_{{\rm FP}}^{2}+12z_{{\rm FP}}^{3}\right)}{32z_{{\rm FP}}\left(1-z_{{\rm FP}}^{2}\right)\left(2z_{{\rm FP}}-3\right)\left(z_{{\rm FP}}-6\right)}.\label{eq:alphak_C60}
\end{eqnarray}
However, here already the expansion around the fixed point excludes
$z_{{\rm FP}}=\pm1$, both being outside the domain of poles, see
Fig.~\ref{fig:DSGpoles}. Inserting these results into $X_{k}$ in
Eq.~(\ref{eq:Xmatrix}) and expanding in powers of $\zeta=z-z_{{\rm FP}}$
yields for some generic component:
\begin{eqnarray}
\left[X_{k}\right]_{11} & \sim & -\zeta^{-1}\frac{\left(z_{{\rm FP}}-6\right)\left(2z_{{\rm FP}}-3\right)\left(5-z_{{\rm FP}}-3z_{{\rm FP}}^{2}\right)\left(3+z_{{\rm FP}}-5z_{{\rm FP}}^{2}\right)}{3\left(3-23z_{{\rm FP}}-41z_{{\rm FP}}^{2}+12z_{{\rm FP}}^{3}\right)^{2}\left({\cal A}\lambda_{1}^{k}\right)}
 -\zeta^{0}\frac{\left(45+72z_{{\rm FP}}-107z_{{\rm FP}}^{2}-24z_{{\rm FP}}^{3}+45z_{{\rm FP}}^{4}\right)}{96z_{{\rm FP}}\left(z_{{\rm FP}}^{2}-1\right)}\nonumber \\
 &  & \quad-\zeta^{1}\left[\frac{5\left(5-z_{{\rm FP}}-3z_{{\rm FP}}^{2}\right)\left(3+z_{{\rm FP}}-5z_{{\rm FP}}^{2}\right)\left(3+7z_{{\rm FP}}-49z_{{\rm FP}}^{2}+12z_{{\rm FP}}^{3}\right)^{2}}{6144z_{{\rm FP}}^{2}\left(z_{{\rm FP}}-6\right)\left(z_{{\rm FP}}^{2}-1\right)^{2}\left(2z_{{\rm FP}}-3\right)}\left({\cal B}\lambda_{2}^{k}\right)\right]+\ldots,\label{eq:X11zeta_C60}\\
 & \sim & \zeta^{-1}O\left(\frac{1}{\lambda_{1}^{k}}\right)+\zeta^{0}O(1)+\zeta^{1}O\left(\lambda_{2}^{k}\right)+\ldots.\nonumber 
\end{eqnarray}
As before, we have to exclude the values $z_{{\rm FP}}=\pm1$. Thus,
we obtain the same result for the scaling in $X_{k}$ as for the Grover
coin in Eq.~(\ref{eq:X11zeta}), further affirming that the universality
class remains unchanged. 

\section{Conclusion\label{sec:Discussion}}

We have explored the properties of the RG applied to discrete-time
(coined) quantum walks for an alternative coin that breaks the reflection
symmetry of the conventional Grover coin. To that end, we have first
explored its effect in the exactly solvable case of such walks on
a \emph{1d}-line. There, we were able to show by explicit computation
that the formal RG fixed-point analysis is generic and unaffected
by the specific value of the Laplace parameter $z=z_{{\rm FP}}$ at
which the analysis is conducted, as long as certain isolated values
are excluded. (We note that also certain \emph{correlated} limits
of the scalar parameters $\vec{a}_{k}$ exist for which non-generic,
specifically, classical random walk results are obtained, as already
indicated in Ref.~\cite{Boettcher13a}. These occur also at isolated
points in $z$ that we will discuss elsewhere and which do not contradict
the generic picture we have presented here.) While the corresponding
classical RG analysis is focused on Laplace poles near the real axis
at $z\to z_{{\rm FP}}=1$, we have shown that it is necessary to consider
fixed points $z_{{\rm FP}}$ anywhere on the complex unit-circle,
as those poles cluster there in distinct domains. Especially in the
case of a non-reflective coin, these domains may in fact exclude the
real axis of the complex $z$-plane. Yet, despite of those variations
in the RG-analysis between distinct coins, the universality classes
of the quantum walks studied here remain unchanged and seem to depend
only on the network that is studied. We have demonstrated these conclusions
for the non-trivial case of the dual Sierpinski gasket (DSG).

\bibliographystyle{apsrev4-1}
\bibliography{/Users/stb/Boettcher}

\end{document}